\documentclass[fleqn,usenatbib]{mnras}

\usepackage{newtxtext,newtxmath}

\usepackage[T1]{fontenc}

\DeclareRobustCommand{\VAN}[3]{#2}
\let\VANthebibliography\thebibliography
\def\thebibliography{\DeclareRobustCommand{\VAN}[3]{##3}\VANthebibliography}

\usepackage{graphicx}	
\usepackage{amsmath}	
\usepackage{multirow}   
\usepackage[utf8]{inputenc} 
\usepackage{tabularx}
\usepackage{comment}
\usepackage{pdflscape}
\usepackage{scalerel}
\usepackage{tikz}

\usetikzlibrary{svg.path}

\newcommand{\plotsidesize}[2]
 {\centering \leavevmode \includegraphics[width={#2\textwidth}]{#1}}

\newcommand{\msun}{M_{\sun}}

\newcommand{\GIZMO}{{\small GIZMO}}
\newcommand{\Hmol}{{\rm H}_{2}}

\newcommand{\qPAH}{q_{\rm PAH}}

\newcommand{\nH}{n_{\rm H}}
\newcommand{\nHn}{n_{\rm H,neutral}}
\newcommand{\NHn}{N_{\rm H,neutral}}
\newcommand{\cmcubed}{ {\rm cm}^{-3} }
\newcommand{\Zcrit}{Z_{\rm crit}}

\newcommand{\Zsol}{{\rm Z}_{\sun}}
\newcommand{\Msol}{{M}_{\sun}}
\newcommand{\Rvir}{R_{\rm vir}}
\newcommand{\Mvir}{M_{\rm vir}}
\newcommand{\OH}{12+{\rm log_{10}(O/H)}}
\newcommand{\teq}{\tau_{\rm equil}}
\newcommand{\tmetal}{\tau_{\rm metal}}

\defcitealias{choban_2022:GalacticDustupModelling}{C22}

\newcommand{\gizmourl}{\href{http://www.tapir.caltech.edu/~phopkins/Site/GIZMO.html}{\url{http://www.tapir.caltech.edu/~phopkins/Site/GIZMO.html}}}

\newcommand{\datastatement}[1]{\begin{small}\section*{Data Availability Statement}\end{small}{\noindent #1}\vspace{5pt}}

\definecolor{orcidlogocol}{HTML}{A6CE39}
\tikzset{
  orcidlogo/.pic={
    \fill[orcidlogocol] svg{M256,128c0,70.7-57.3,128-128,128C57.3,256,0,198.7,0,128C0,57.3,57.3,0,128,0C198.7,0,256,57.3,256,128z};
    \fill[white] svg{M86.3,186.2H70.9V79.1h15.4v48.4V186.2z}
                 svg{M108.9,79.1h41.6c39.6,0,57,28.3,57,53.6c0,27.5-21.5,53.6-56.8,53.6h-41.8V79.1z M124.3,172.4h24.5c34.9,0,42.9-26.5,42.9-39.7c0-21.5-13.7-39.7-43.7-39.7h-23.7V172.4z}
                 svg{M88.7,56.8c0,5.5-4.5,10.1-10.1,10.1c-5.6,0-10.1-4.6-10.1-10.1c0-5.6,4.5-10.1,10.1-10.1C84.2,46.7,88.7,51.3,88.7,56.8z};
  }
}
\newcommand\orcidicon[1]{\href{https://orcid.org/#1}{\mbox{\scalerel*{
\begin{tikzpicture}[yscale=-1,transform shape]
\pic{orcidlogo};
\end{tikzpicture}
}{|}}}}

\title[Evolution of Galactic Dust Populations]{A Dusty Locale: Evolution of Galactic Dust Populations from Milky Way to Dwarf-Mass Galaxies}
\author[C. R. Choban et al.]{
\parbox[t]{\textwidth}{
        Caleb R. Choban\orcidicon{0000-0001-9200-169X}$^{1,2}$\thanks{email: cchoban@iu.edu},
        Du\v{s}an Kere\v{s}\orcidicon{0000-0002-1666-7067}$^{1}$,
        Karin M. Sandstrom\orcidicon{0000-0002-4378-8534}$^{1}$,
        Philip F. Hopkins\orcidicon{0000-0003-3729-1684}$^{3}$,
        Christopher C. Hayward\orcidicon{0000-0003-4073-3236}$^{4}$
        and Claude-Andr{\'e} Faucher-Gigu{\`e}re\orcidicon{0000-0002-4900-6628}$^{5}$
} \vspace*{4pt} \\
$^{1}$ Department of Astronomy \& Astrophysics and Department of Physics, Center for Astrophysics and Space Science, University of California at San Diego, \\ La Jolla, CA 92093, USA \\
$^{2}$ Department of Astronomy, Indiana University, Bloomington, IN 47405, USA \\
$^{3}$ TAPIR, Mailcode 350-17, California Institute of Technology, Pasadena, CA 91125, USA \\
$^{4}$ Center for Computational Astrophysics, Flatiron Institute, New York, NY 10010 USA \\
$^{5}$ Department of Physics and Astronomy and CIERA, Northwestern University, Evanston, IL 60208, USA \\
}

\date{Accepted XXX. Received YYY; in original form ZZZ}

\pubyear{2023}

\begin{document}
\label{firstpage}
\pagerange{\pageref{firstpage}--\pageref{lastpage}}
\maketitle

\begin{abstract}
Observations indicate dust populations vary between galaxies and within them, suggesting a complex life cycle and evolutionary history.
Here we investigate the evolution of galactic dust populations across cosmic time using a suite of cosmological zoom-in simulations from the Feedback in Realistic Environments (FIRE) project, spanning $M_{\rm vir}=10^{9-12}\Msol;\,M_{*}=10^{6-11}\,\Msol$.
Our simulations incorporate a dust evolution model that accounts for the dominant sources of dust production, growth, and destruction and follows the evolution of specific dust species. 
All galactic dust populations in our suite exhibit similar evolutionary histories, with gas-dust accretion being the dominant producer of dust mass for all but the most metal-poor galaxies. Similar to previous works, we find the onset of efficient gas-dust accretion occurs above a `critical' metallicity threshold ($\Zcrit$).
Due to this threshold, our simulations reproduce observed trends between galactic D/Z and metallicity and element depletion trends in the ISM. 
However, we find $\Zcrit$ varies between dust species due to differences in key element abundances, dust physical properties, and life cycle processes resulting in $\Zcrit\sim0.05\Zsol,\,0.2\Zsol,\,0.5\Zsol$ for metallic iron, silicates, and carbonaceous dust, respectively. 
These variations could explain the lack of small carbonaceous grains observed in the Magellanic Clouds.
We also find a delay between the onset of gas-dust accretion and when a dust population reaches equilibrium, which we call the equilibrium timescale ($\teq$). 
The relation between $\teq$ and the metal enrichment timescale of a galaxy, determined by its recent evolutionary history, can contribute to the scatter in the observed relation between galactic D/Z and metallicity.
\end{abstract}




\section{Introduction}

Observations of the Milky Way (MW) and the Large and Small Magellanic Clouds (LMC \& SMC) reveal significant variations in their respective dust populations. 
The 2175 \r{A} feature/bump in dust extinction curves, which correlates with the abundance of small carbonaceous grains\footnote{While it is currently unknown whether PAHs or small carbonaceous grains (i.e. amorphous graphite) are the dominant carrier for the 2175 \r{A} feature, their life cycles could be intimately linked. One proposed formation mechanism for PAHs is through photoprocessing, where hydrogen atoms are removed from small carbonaceous dust grains \citep[i.e. aromatization;][]{rau_2019:ModellingEvolutionPAH,hirashita_2020:SelfconsistentModellingAromatic}.
However, other possible formation mechanisms via AGB winds \citep{galliano_2008:StellarEvolutionaryEffects} or directly through low-temperature chemical channels in dense molecular clouds \citep{parker_2012:LowTemperatureFormation} have not been ruled out.}, shows a strong decrease between the Milky Way and LMC, and an almost complete lack of the feature for the SMC \citep[e.g.][]{pei_1992:InterstellarDustMilky, weingartner_2001:DustGrainSizeDistributions}.
Mid-infrared dust emission produced by polycyclic aromatic hydrocarbons (PAHs) shows a dramatic decrease in the average fraction of dust mass comprised of PAHs ($\qPAH$; \citealt{draine_2007:InfraredEmissionInterstellar}) from the Milky Way (${\sim}4.6\%$) to the LMC (${\sim}3.3\%$) and SMC (${\sim}1.0\%$) \citep{li_2001:InfraredEmissionInterstellar,weingartner_2001:DustGrainSizeDistributions,chastenet_2019:PolycyclicAromaticHydrocarbon}. 
UV-based observations of gas-phase element depletions within the MW, LMC, and SMC indicate that the fraction of metals locked in dust (dust-to-metals ratio; D/Z) increases with gas surface density \citep{jenkins_2009:UnifiedRepresentationGasPhase,jenkins_2017:InterstellarGasphaseElement,roman-duval_2022:METALMetalEvolutiona}. These observations also reveal varying galactic D/Z, corresponding with each galaxy's relative metallicities, and considerable variation in dust chemical composition. However, the lack of observed C and O depletions in the LMC and SMC results in an incomplete picture.

Looking further afield, observations of local galaxies tell a similar story. Dust emission surveys find an overall increase of galaxy-integrated D/Z with metallicity along with a large (>1 dex) scatter at any given metallicity, but the exact relation varies between studies \citep{remy-ruyer_2014:GastodustMassRatios,devis_2019:SystematicMetallicityStudy}. These studies also find a dependence of $\qPAH$ with logarithmic metallicity, with a threshold galactic metallicity of $12+{\rm log_{10}(O/H)}\sim8.0$ above which $\qPAH$ rapidly increases \citep{draine_2007:DustMassesPAH,remy-ruyer_2015:LinkingDustEmission,aniano_2020:ModelingDustStarlight}.
Local and high-z dust attenuation curves in galaxies exhibit a wide range of 2175 \r{A} feature strengths \citep[e.g.][]{salim_2020:DustAttenuationLaw,shivaei_2022:UV2175AAttenuation}. 
However, the strength/absence of this feature is not solely determined by the underlying dust population-derived extinction curve. In particular, radiative transfer effects caused by interstellar medium (ISM) clumpiness, star-dust geometry, and adopted dust albedo can modify the expected attenuation feature \citep{granato_2000:InfraredSideGalaxy, panuzzo_2007:UltravioletDustAttenuation,seon_2016:RadiativeTransferModel,narayanan_2018:TheoryVariationDust}.
Observations of damped Ly$\alpha$ systems (DLAs) show increasing gas-phase element depletions with metallicity and no clear relation with redshift \citep{peroux_2020:CosmicBaryonMetal,roman-duval_2022:METALMetalEvolution}.
These observations suggest an evolving dust population in both amount and chemical composition, with a decreasing total fraction of metals locked in dust and a scarcity of carbonaceous dust in low-metallicity environments. 
The exact cause of this is unknown, but detailed dust evolution models that account for the main processes of the dust life cycle coupled with galaxy formation models can help elucidate this issue.

The current dust life cycle paradigm posits all dust grains begin their life in the ejecta of Type II supernovae (SNe II) and stellar winds of asymptotic giant branch (AGB) stars, where a fraction of metals within these winds coalesce into dust \citep{draine_1990:EvolutionInterstellarDust,todini_2001:DustFormationPrimordial,dwek_2005:InterstellarDustWhat,ferrarotti_2006:CompositionQuantitiesDust}. The grains are then primarily destroyed by SNe shocks. In isolation, stellar dust production alone cannot explain the dust content of either the Milky Way \citep[e.g.][]{mckee_1989:DustDestructionInterstellar,draine_2009:InterstellarDustModels} or LMC \citep{zhukovska_2013:DustInputAGB}, requiring another growth/production mechanism. 
Observations of gas-phase element depletions \citep{jenkins_2009:UnifiedRepresentationGasPhase,roman-duval_2021:METALMetalEvolution} and dust emission \citep{roman-duval_2014:DustGasMagellanic,roman-duval_2017:DustAbundanceVariations,chiang_2018:SpatiallyResolvedDusttometals,clark_2023:QuestMissingDust} show D/Z increases from diffuse to dense regions of the Milky Way and Local Group galaxies. These provide strong evidence that dust grows via the accretion of gas-phase metals in dense environments and could be the main producer of dust mass within these galaxies.
However, the exact details of the accretion process are poorly understood, and many questions have yet to be completely answered. {\it When does gas-dust accretion become efficient, and does it differ between dust species? How long does it take for accretion to build up a sizable dust mass within a galaxy? Can accretion explain dusty galaxies at high redshift?}

Analytical galactic dust evolution models were first developed to answer these questions, giving rise to the concept of a `critical' metallicity ($\Zcrit$) threshold \citep{inoue_2011:OriginDustGalaxies,asano_2013:DustFormationHistory,zhukovska_2014:DustOriginLatetype,feldmann_2015:EquilibriumViewDust,popping_2017:DustContentGalaxies,triani_2020:OriginDustGalaxies}. While the exact definition varies between works, they all agree there is a `critical' metallicity above which gas-dust accretion becomes the dominant source of dust mass, rapidly increasing the expected galactic D/Z\footnote{These findings are not unanimous among all works. In particular, \citet{priestley_2022:ImpactMetallicitydependentDust} suggests the contribution of dust growth via accretion may be overestimated if both high stardust creation efficiencies and increased SNe dust destruction in low-metallicity environments are assumed.}. \citet{inoue_2011:OriginDustGalaxies} found that $\Zcrit$ is determined by the competition between growth via accretion and destruction via SNe shocks. However, equilibrium models developed by \citet{feldmann_2015:EquilibriumViewDust} suggest the dilution of dust via dust-poor gas inflows dominates over destruction via SNe shocks when determining $\Zcrit$, especially for dwarf galaxies. \citet{zhukovska_2008:EvolutionInterstellarDust} developed analytical models that track the evolution of specific dust species and found that for a Milky Way-mass galaxy, silicates, carbonaceous, and metallic iron dust species have different $\Zcrit$.
In later works, this model was used to investigate the evolution of dust within late-type dwarf galaxies experiencing episodic starbursts \citep{zhukovska_2014:DustOriginLatetype}. 
They concluded that such galaxies have lower $\Zcrit$ than galaxies with constant star formation due to long quiescent periods between starbursts where dust has ample time to grow.

More recently, numerical dust evolution models integrated into galaxy formation and evolution simulations have been put to the task 
\citep[e.g.][]{bekki_2015:CosmicEvolutionDust,mckinnon_2016:DustFormationMilky,zhukovska_2016:ModelingDustEvolution,mckinnon_2017:SimulatingDustContent,aoyama_2020:GalaxySimulationEvolution,granato_2021:DustEvolutionZoomin}.
The dust evolution models utilized in these works differ in their methodology and included physics
and the galaxy simulations employed primarily encompass cosmological volumes, relying on varying sub-resolution star formation, feedback, and ISM prescriptions. Despite these differences, they all agree with the `critical' metallicity concept (albeit with a large range $\Zcrit \sim 0.03-0.2 \,\Zsol$) above which the average galactic D/Z rapidly increases due to dust growth via accretion \citep{hou_2019:DustScalingRelations,li_2019:DusttogasDusttometalRatio,graziani_2020:AssemblyDustyGalaxies,parente_2022:DustEvolutionMUPPI}.
A majority of these works also find that $\Zcrit$ and the resulting relation between D/Z and Z have little evolution over redshift\footnote{We point the reader to \citet{popping_2022:ObservedCosmicEvolution} for a direct comparison of the predicted relations between D/Z and Z and its evolution over redshift for a selection of analytical and numerical dust models.}.
However, the exact cause of the large scatter in observed D/Z at any given Z is debated. 
\citet{parente_2022:DustEvolutionMUPPI} predicts this is due to the fraction of cold gas in each galaxy, with higher cold gas fractions allowing for the faster build-up of dust via accretion. 
\citet{li_2019:DusttogasDusttometalRatio} proposes a more complex dependence on galactic metal enrichment history reflected by
$Z$ and $M_*$ and evolutionary stages quantified by gas content and depletion timescales. \citet{li_2021:OriginDustExtinction} adds that variance in galactic mass-averaged grain sizes (and thus gas-dust accretion rates) can also drive this scatter.
Regarding the differences in the evolution between silicate and carbonaceous dust species, \citet{granato_2021:DustEvolutionZoomin} and \citet{parente_2022:DustEvolutionMUPPI} predict that silicate growth lags behind carbonaceous growth due to the overall lower abundance of Si compared to C, with silicate dust only dominating the dust mass past $\Zcrit\gtrsim0.15 \Zsol$, $M_{*} \gtrsim 10^{8.5} \Msol$.

While these dust evolution models generally agree in terms of galaxy-integrated dust observations, they have three major limitations that make them ill-suited tools for investigating the gas-dust accretion process.
First, models integrated into cosmological galaxy simulations do not resolve the multi-phase ISM, which is necessary to accurately model each process in the dust life cycle. 
Second, most models assume a single, chemically ambiguous dust population, affecting how gas-dust accretion is modeled. 
Only \citet{granato_2021:DustEvolutionZoomin} and \citet{parente_2022:DustEvolutionMUPPI} are the exception, tracking the evolution of chemically distinct dust species\footnote{While other dust evolution models which track the evolution of separate dust species exist \citep[e.g.][]{hirashita_2019:RemodellingEvolutionGrain,narayanan_2023:FrameworkModelingPolycyclic}, they are not yet incorporated into cosmological simulations.}.
Third, most models assume a constant grain size distribution (typically a MRN size distribution $ dn_{\rm gr}(a)/da \propto a^{-3.5}$; \citealt{mathis_1977:SizeDistributionInterstellar}), which affects the predictions of most dust processes due to their dependence on a dust population's effective grain size. 
The only exceptions to this are \citet{li_2021:OriginDustExtinction}, who model multiple grain size bins, and \citet{granato_2021:DustEvolutionZoomin} and \citet{parente_2022:DustEvolutionMUPPI}, who utilize a two-size approximation \citep{hirashita_2015:TwosizeApproximationSimple}.
\citet[][\citetalias{choban_2022:GalacticDustupModelling} hereafter]{choban_2022:GalacticDustupModelling} developed two separate dust evolution models to investigate the effects of the first and second assumptions.
They found that only a model that tracks the evolution of chemically-distinct dust species and incorporates a physically motivated, two-phase dust growth routine can reproduce the observed scaling relation between individual element depletions and D/Z with column density and local gas density in the Milky Way. This relation results from the equilibrium between SNe dust destruction and dust growth via accretion. Therefore, such a model is crucial for investigating the evolution of galactic dust populations and testing the importance of gas-dust accretion in that evolution. 
However, \citetalias{choban_2022:GalacticDustupModelling} only utilized simulations of an idealized, non-cosmological Milky Way-like galaxy. Due to their idealized nature, these simulations do not capture the formation and hierarchical growth of galaxies in fully cosmological settings across cosmic time and do not include a realistic baryon cycle in the IGM/CGM/ISM. Consequently, these predictions may not hold for the entire evolutionary history of a galaxy or lower-mass galaxies.

In this work, we present a subset of cosmological zoom-in simulations of Milky Way to dwarf-halo mass galaxies from the Feedback in Realistic Environments (FIRE) project\footnote{\url{http://fire.northwestern.edu}} rerun with the integrated ``Species'' dust evolution model presented in \citetalias{choban_2022:GalacticDustupModelling}. 
This model tracks the evolution of specific dust species with set chemical compositions and incorporates a physically motivated dust growth routine. 
We highlight the primary limitation of this model is the exclusion of evolving grain sizes.
With these simulations, we investigate how galactic dust populations evolve in both their amount and chemical composition, focusing on the determining factors for when gas-dust accretion dominates and how it differs between dust species.
We find that gas-dust accretion is the dominant producer of dust mass for all but the most metal-poor galaxies and, in the case of the MW, dominates for the majority of the galaxy's life.
We discover that the onset of rapid growth via gas-dust accretion differs between dust species, arising from differences in their key element abundances, physical properties, and life cycle processes. These differences can explain the variable dust population, in both amount and composition, in the MW, LMC, and SMC. We also find a delay between the onset of rapid dust growth via accretion and when a dust population reaches equilibrium between growth and destruction processes. The relation between this delay and the metal enrichment timescale of a galaxy can contribute to the scatter in observed D/Z at any given metallicity.

This paper is organized as follows. In Section~\ref{sec:methods}, we provide a brief overview of our simulation sample along with the galaxy formation and dust evolution model used. In Section~\ref{sec:results}, we present the results of our simulations, focusing on the evolution of the galactic and dust population properties for each galaxy in Section~\ref{sec:cosmic_evolution} and comparing them with local observations in Section~\ref{sec:element_depletions} \&~\ref{sec:dust_emission}. We discuss the concept of a `critical' metallicity threshold, which marks the onset of efficient gas-dust accretion, in Section~\ref{sec:efficient_accretion} and an `equilibrium' timescale, which is the time over which a dust population's mass builds up via accretion to an effective equilibrium, in Section~\ref{sec:dust_buildup}. Finally, we present our conclusions in Section~\ref{sec:conclusions}.

\section{Methodology} \label{sec:methods}

\begin{table*}
	\centering
	\begin{tabular}{cccccccc} 
		\hline
		Name & $\Mvir \; (\msun)$ & $\Rvir$ (kpc) & $M_{*} \; (\msun)$ & $R_{1/2}$ (kpc) & Resolution ($\msun$) & Notes\\
		\hline
		m12i & 9.5E11 & 203 & 7.2E10 & 2.9 & 7100 & MW-mass spiral w/ minor mergers $z\lesssim0.7$ \& recent flybys\\
		m11v\_halo0 & 2.6E11 & 131 & 1.6E9 & 4.13 & 7100 & LMC-mass spiral which evolves in isolation\\
            m11d & 2.4E11 & 131 & 3.9E9 & 2.77 & 7100 & LMC-mass dwarf w/ several mergers before z$\sim1$ \\
            m11e & 1.3E11 & 107 & 1.4E9 & 1.89 & 7100 & LMC-mass dwarf with recent major merger at z$\sim$0.1 \\
            m11v\_halo2  & 5.3E10 & 34.4 & 2.1E9 & 3.1 & 7100 & LMC-mass dwarf which evolves in isolation\\
            m11i & 6.0E10 & 82.4 & 1.9E8 & 2.32 & 7100 & bursty SMC-mass dwarf w/ major merger at $z\sim0.8-0.5$\\
		m10q & 6.6E9 & 38.6 & 1.8E6 & 0.49 & 500 & low-mass dwarf w/ major blowout at $z\sim0.3$\\
	   \hline
	\end{tabular}
	\caption{Parameters describing initial conditions of simulations in this paper. The m11v simulation is unique since there are two galaxies (labeled halo0 and halo2) about to merge at simulation end ($\sim$70 kpc away), with halo2's dark matter halo being a subhalo of halo0. 
    \textbf{(1)} Name of simulation. 
    \textbf{(2)} Virial mass of dark matter halo at $z=0$. 
    \textbf{(3)} Virial radius of dark matter halo at $z=0$. 
    \textbf{(4)} Stellar mass within $\leq 3 R_{1/2}$ of the galactic center at $z=0$. 
    \textbf{(5)} Stellar half-mass radius at z$=0$. 
    \textbf{(6)} Mass of gas and star particles. 
    \textbf{(7)} Additional notes for each simulation.}
    \label{tab:simulations}
\end{table*}

To study the evolution of galactic dust populations within Milky Way to dwarf-mass galaxies, we reran a subset of cosmological simulations from the FIRE suite presented in \citet{hopkins_2018:FIRE2SimulationsPhysics} and \citet{el-badry_2018:GasKinematicsMorphology}.
We selected 7 galaxies with a broad range of stellar ($10^6 \, \Msol < M_{*} < 10^{11} \, \Msol$) and halo ($10^{9} \, \Msol < M_{\rm vir} < 10^{12} \, \Msol$) masses at present day. We give the exact details of the $z=0$ halo virial mass, virial radius, stellar mass, stellar half-mass radius, and mass resolution for each simulation in Table~\ref{tab:simulations}.

All simulations in this work are run with the \GIZMO\ code base \citep{hopkins_2015:NewClassAccurate} in the meshless finite-mass (MFM) mode with FIRE-2 \citep{hopkins_2018:FIRE2SimulationsPhysics} model of star formation and stellar feedback (see Appendix~\ref{Appendix_FIRE3} for a subset of simulations run with FIRE-3; \citealt{hopkins_2023:FIRE3UpdatedStellar}). FIRE-2, an updated version of FIRE \citep{hopkins_2014:GalaxiesFIREFeedback}, incorporates multiple sources of stellar feedback, specifically stellar winds (O/B and AGB), ionizing photons, radiation pressure, and supernovae (both Types Ia and II). Gas cooling is followed for T = 10 - $10^{10}$ K including free-free, Compton, metal-line, molecular, fine-structure, and dust collisional processes while gas is also heated by cosmic rays, photo-electric, and photoionization heating by both local sources and a uniform but redshift dependent meta-galactic background \citep{faucher-giguere_2009:NewCalculationIonizing}, including the effect of self-shielding. Star formation is only allowed in cold, molecular, and locally self-gravitating regions with number densities above $\nH = 1000 \, {\rm cm}^{-3}$.

Each star particle represents a stellar population with a known mass, age, and metallicity assuming a \citet{kroupa_2002:InitialMassFunction} initial mass function (IMF) from $0.1-100\; \rm \msun$. The luminosity, mass loss rates, and SNe II rates of each star particle are calculated based on the {\small STARBURST99} \citep{leitherer_1999:Starburst99SynthesisModels} libraries, and SNe Ia rates following \citet{mannucci_2006:TwoPopulationsProgenitors}. Metal yields from SNe II, Ia, and AGB winds are taken from \citet{nomoto_2006:NucleosynthesisYieldsCorecollapse}, \citet{iwamoto_1999:NucleosynthesisChandrasekharMass}, and \citet{izzard_2004:NewSyntheticModel} respectively. Evolution of eleven species (H, He, C, N, O, Ne, Mg, Si, S, Ca, and Fe) is tracked for each gas cell. Sub-resolution turbulent metal diffusion is modeled as described in \citet{su_2017:FeedbackFirstSurprisingly} and \citet{escala_2018:ModellingChemicalAbundance}. FIRE-2 adopts the older \citet{anders_1989:AbundancesElementsMeteoritic} solar metal abundances with $Z\sim 0.02$ so whenever we mention solar abundances we are referring to the Andres \& Gravesse abundances.\footnote{In Appendix~\ref{Appendix_FIRE3} we compare a subset of simulations run with FIRE-2 and FIRE-3. While FIRE-3 incorporates numerous improvements, including the adoption of the newer \citet{asplund_2009:ChemicalCompositionSun} proto-solar abundances with $\Zsol\sim0.014$ and updated nucleosynthesis yields, we find no major differences in our findings between the two versions.}

FIRE is ideally suited to investigate galactic dust evolution over cosmic time given its success in matching a wide range of observations related to galaxies, including the mass-metallicity relation and its evolution over redshift (\citealt{ma_2016:OriginEvolutionGalaxy}; Marszewski et al. in prep.) and the Kennicutt–Schmidt star formation law \citep{hopkins_2014:GalaxiesFIREFeedback,orr_2018:WhatFIREsStar,gurvich_2020:PressureBalanceMultiphase}. This success is owed to the high resolution, star formation criteria, cooling to low temperatures, and multi-channel stellar feedback of FIRE, all of which result in a reasonable ISM phase structure and giant molecular cloud (GMC) mass function \citep{benincasa_2020:LiveFastYoung}. These also lead to the self-consistent development of galactic winds that eject large amounts of gas \citep{muratov_2015:GustyGaseousFlows, angles-alcazar_2017:CosmicBaryonCycle} and metals \citep{muratov_2017:MetalFlowsCircumgalactic, hafen_2019:OriginsCircumgalacticMedium,pandya_2021:CharacterizingMassMomentum} out of galaxies, preventing excessive star formation and leading to a plausible stellar-halo mass relation.

Our simulations utilize the integrated ``Species'' dust evolution model presented in \citetalias{choban_2022:GalacticDustupModelling}, which we refer to the reader for full details. This model includes the dominant sources of dust production, tracking and differentiating between dust created from SNe Ia and II, AGB stars, and dust growth from gas-phase metal accretion in the ISM. It includes the dominant dust destruction mechanisms, accounting for dust destroyed by SNe shocks, thermal sputtering, and astration (dust destroyed during the formation of stars). These processes are modeled self-consistently, owing to the FIRE model's in-depth treatment of the multi-phase ISM and capacity to time-resolve individual SNe events \citep{hopkins_2018:HowModelSupernovae}. Notably, we restrict gas-dust accretion to cool ($T\leq300$ K) gas and destroy dust locally around individual SNe events, allowing us to track the local variability of dust in the ISM.
We also follow the evolution of specific dust species (carbonaceous, silicates, and silicon carbide) and theoretical oxygen-bearing (O-reservoir) and nanoparticle metallic iron (Nano-iron) dust species with set chemical compositions. Consequently, this means each dust species has a key element\footnote{Here key element refers to the element for which $n/i$ has the lowest value, where $n$ is the number abundance of the element and $i$ is the number of atoms of the element in one formula unit of the dust species under consideration.} that limits their individual accretion growth rates and the maximum formable amount of said dust species. 
We also incorporate sub-resolution turbulent dust diffusion, which follows the metal diffusion prescription in FIRE, and a dense molecular gas scheme. 
This scheme is critical to account for Coulomb enhancement of gas-dust accretion in atomic/diffuse molecular gas and the reduction in carbonaceous dust accretion due to the lock-up of gas-phase C into CO in dense molecular gas.

We highlight that all cooling and heating processes and radiative transfer modeled in our simulations are not coupled with our dust evolution model and instead follow the default assumptions in FIRE-2. Specifically, dust heating and cooling and radiative transfer assume a constant D/Z ratio, and metal-line cooling assumes no metals are locked in dust. 
This choice was initially made to avoid possible changes to galaxy evolution and resulting $z=0$ galactic properties, which could affect our dust evolution predictions.
However, in Appendix~\ref{Appendix_FIRE3} we show that our predictions for dust population evolution hold even for large changes in galaxy evolution due to differing ISM physics and star formation when using FIRE-3.
In future work, we will investigate what effects the full integration of our dust evolution model with FIRE physics has on predicted galaxy evolution.

\section{Results} \label{sec:results}

\begin{figure*}
    \plotsidesize{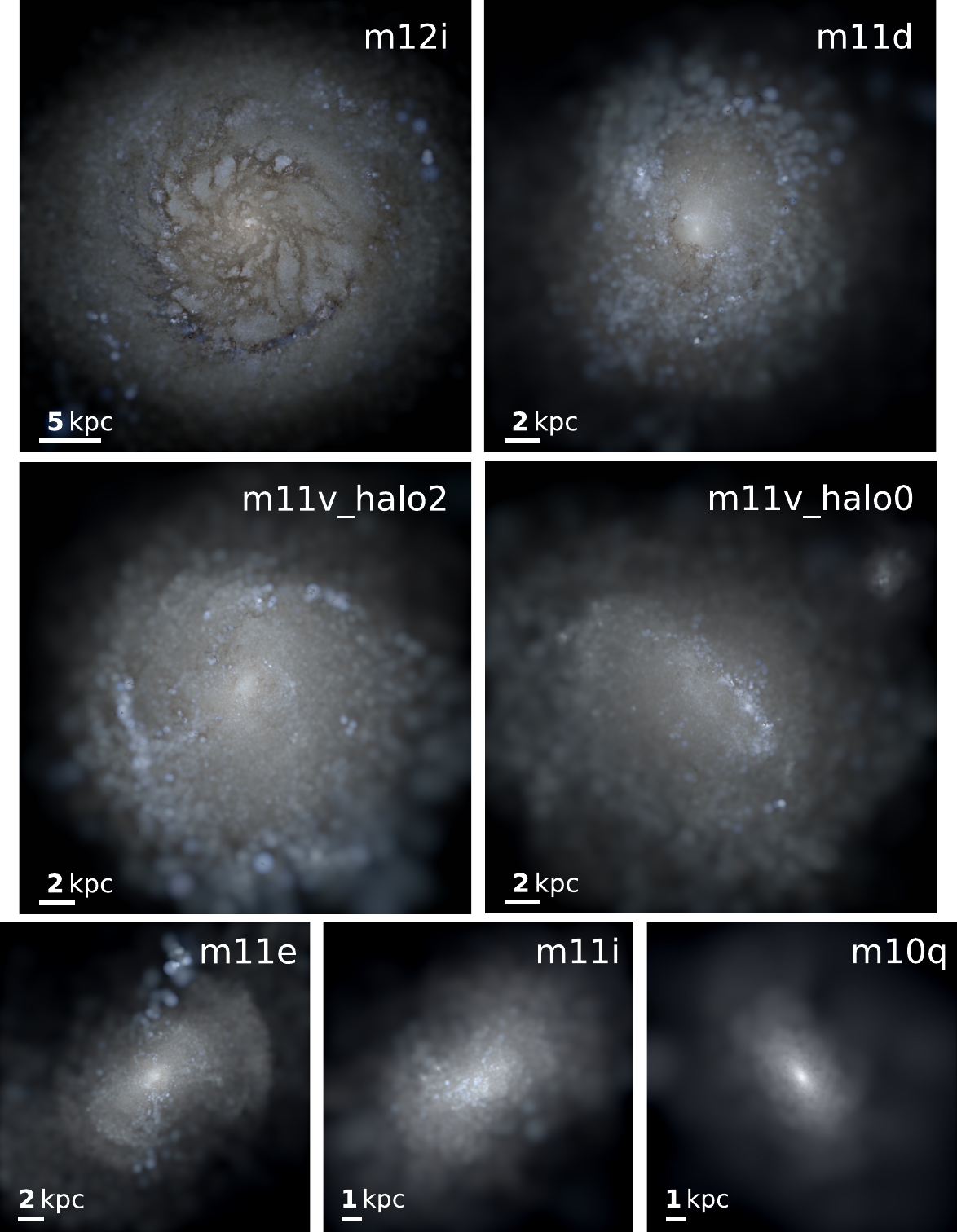}{0.9}
    \vspace{-0.25cm}
    \caption{Face-on images of our spiral and dwarf galaxies at $z=0$ (note the differences in scale). Each image is a mock Hubble Space Telescope {\it ugr} composite accounting for dust attenuation utilizing the D/Z produced by our model along with assumed MW, LMC, or SMC dust opacities depending on the median gas cell Sil-to-C ratio. 
    \textbf{m12i}: A compact, Milky Way-mass spiral galaxy that experienced multiple minor mergers during its life. \textbf{m11d}: An LMC-mass galaxy with bursty star formation and multiple mergers throughout its life. \textbf{m11v\_halo2}: An LMC-mass galaxy with faint spiral arms that evolved in isolation for a majority of its life. \textbf{m11v\_halo0}: An LMC-mass galaxy that evolved in isolation for a majority of its life and experienced a recent burst in star formation. \textbf{m11e}: An irregular dwarf galaxy that spent most of its life as a compact dwarf, similar to the SMC in mass, until it experienced a recent major merger producing prominent stellar shells. \textbf{m11i}: A sub-SMC-mass galaxy that experienced multiple merger events and a bursty star formation history that has successively stripped gas from the galaxy. \textbf{m10q}: A low-mass dwarf galaxy that evolved in isolation and experienced a major blowout event.}
    \label{fig:mock_hubble_all}
\end{figure*}

We first showcase mock HST {\it ugr} composite images of each galaxy to highlight the breadth of galaxy types and their varying dust structure contained in our simulation suite.
Fig.~\ref{fig:mock_hubble_all} shows face-on images of all galaxies at $z=0$. These images use STARBURST99 \citep{leitherer_1999:Starburst99SynthesisModels} to compute the stellar spectra for each star particle given their age and metallicity. These are then ray-traced through the ISM using the tracked D/Z for each gas cell produced by our dust model along with assumed MW, LMC, or SMC dust opacities from \citet{pei_1992:InterstellarDustMilky} depending on the median gas cell silicate-to-carbonaceous dust mass (Sil-to-C) ratio\footnote{We define this ratio as Sil-to-C $= (M_{\rm sil} {+} M_{\rm iron} {+} M_{\rm O-res})/M_{\rm carb}$ in our simulations.} in each galaxy (MW$\sim$3, LMC$\sim$5., SMC$\sim$10).
We then volume-render the observed images in each band and construct a {\it ugr} composite image as seen by a distant observer. {\bf m12i}, {\bf m11d}, {\bf m11v\_halo2}, {\bf m11v\_halo0}, and {\bf m11e} show prominent and detailed dust structure to varying degrees, with dark patches produced by the attenuation of UV/optical light from dust tracing the dense gas in each galaxy. {\bf m11i} and {\bf m10q} have no distinguishable features produced by dust due to extremely low D/Z for {\bf m11i} and the complete lack of both dust and gas for {\bf m10q}.

We present the evolution of each galaxy in terms of galactic properties and their dust population, in both composition and amount, in Sec.~\ref{sec:cosmic_evolution}. We then compare our simulations to present-day observations, focusing on gas-phase element depletions in Sec.~\ref{sec:element_depletions} and galaxy-integrated and spatially-resolved IR dust emission in Sec.~\ref{sec:dust_emission}.

\subsection{Cosmic Evolution} \label{sec:cosmic_evolution}

\subsubsection{Galactic Properties \& Dust Population}

\begin{figure*}
    \plotsidesize{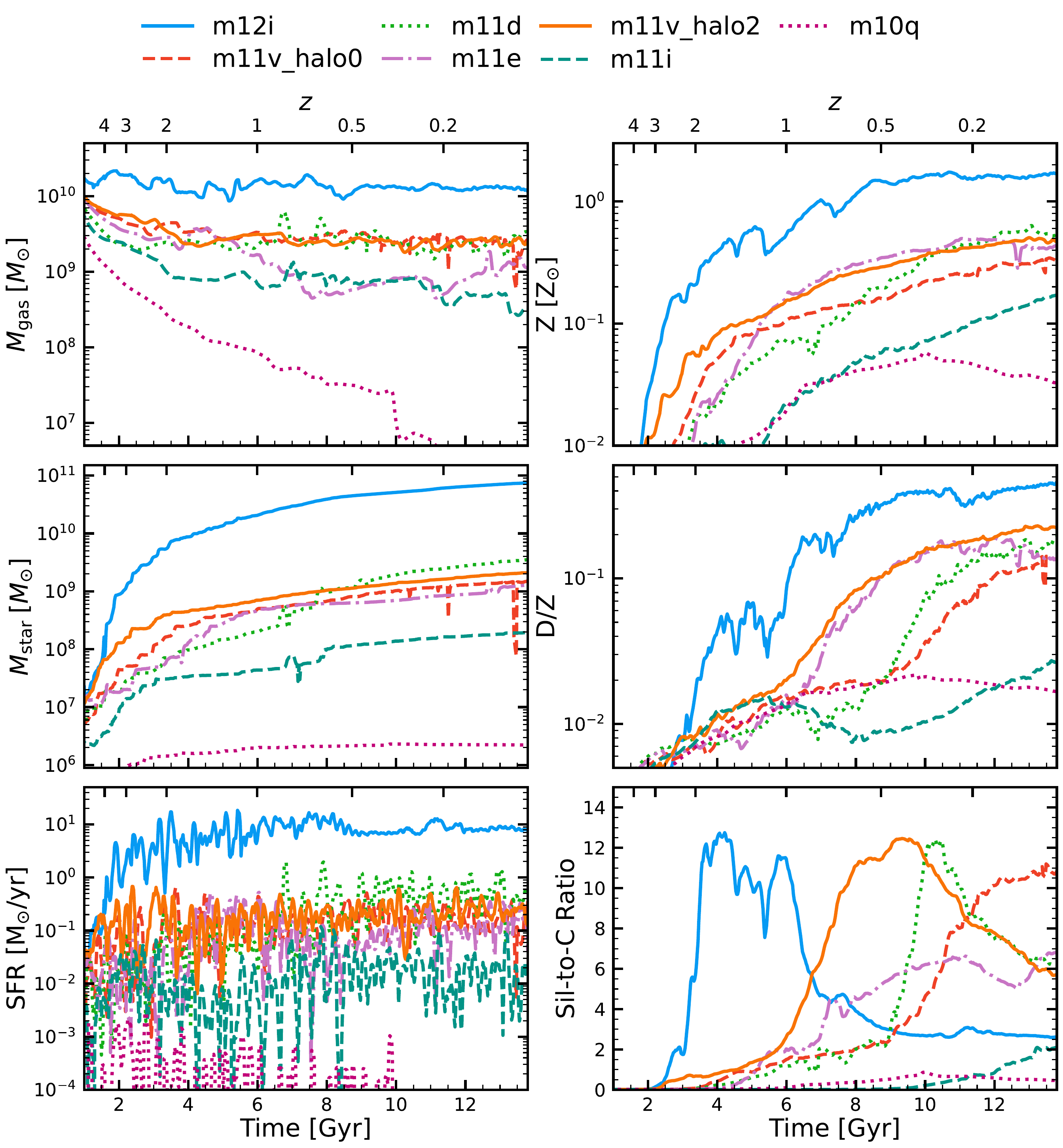}{0.9}
    \vspace{-0.25cm}
    \caption{Evolution of galactic gas, stellar, and dust properties for our simulated galaxies for all gas/stars within 10 kpc of the galactic center. Specifically, total gas mass ({\it top left}), median gas metallicity ({\it top right}), total stellar mass ({\it middle left}), median D/Z ratio ({\it middle right}), star formation rate averaged in 10 Myr intervals ({\it bottom left}), and median silicate-to-carbonaceous dust mass ratio ({\it bottom right}). 
    Our simulation suite covers a wide range of evolutionary histories, star formation rates, and metallicities. However, the resulting dust population evolution is relatively similar for all galaxies. Initially, the dust population is dominated by carbonaceous dust produced by SNe II, and later AGB stars, which results in a low D/Z $\sim0.01$ and Sil-to-C $\sim0$. Eventually, gas-dust accretion becomes efficient for first metallic iron dust, then silicates, and finally carbonaceous dust. The largest increase in D/Z occurs when silicate dust grows efficiently, as can be seen by the rapid increase in Sil-to-C due to silicates dominating the dust mass. When carbonaceous dust begins to grow efficiently, Sil-to-C correspondingly decreases and eventually settles at the typical MW value of $\sim$3. Note that {\bf m10q} experiences a complete blowout event at $\sim$10 Gyr, evacuating almost all gas and dust from the galaxy.
    }
    \label{fig:all_galaxy_evolution}
\end{figure*}

In Fig.~\ref{fig:all_galaxy_evolution}, we show the evolution of various properties for each galaxy, specifically total gas mass, stellar mass, star formation rate (SFR), median gas cell\footnote{Whenever we mention a median value for gas cells, we specifically mean the mass-weighted median. While these values can differ due to deviations in individual gas parcel mass from the stated resolutions in Table~\ref{tab:simulations}, caused by SNe ejecta and stellar wind mass injection, these deviations are minor.} metallicity, median D/Z, and median Sil-to-C ratio. These values are determined from star particles and gas cells within 10 kpc of the galactic center for all galaxies\footnote{We avoid using an evolving radius, such as  $\Rvir$, as our cutoff when determining galactic properties in order to avoid jumps due to merger events. A radius of 10 kpc encompasses the ISM of each galaxy while not including a sizable fraction of the galactic halo, which is sufficient for our focus on interstellar dust.}.
We provide a brief description of each galaxy's evolutionary history, noting any merger events, in Table~\ref{tab:simulations}.

\begin{figure*}
    \plotsidesize{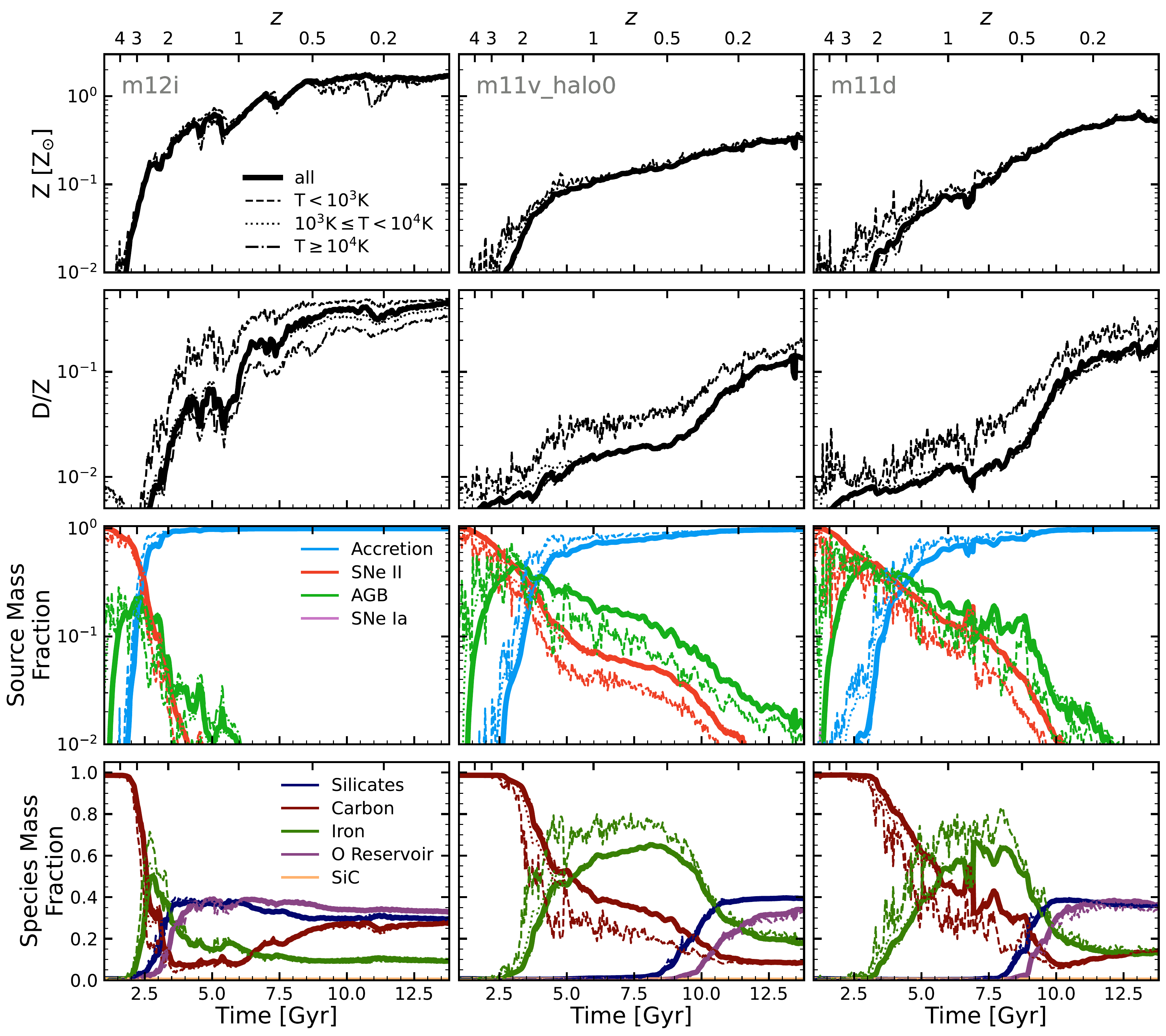}{0.99}
    \vspace{-0.25cm}
    \caption{Time evolution of median metallicity (\textit{top}), D/Z (\textit{second from top}), dust creation source mass fraction (\textit{second from bottom}), and dust species mass fraction (\textit{bottom}) for {\bf m12i} (\textit{left}), {\bf m11v\_halo0} (\textit{middle}), and {\bf m11d} (\textit{right}). Median values are given for all gas within 10 kpc of the galactic halo (\textit{solid}), cold neutral gas (\textit{dashed}), warm neutral gas (\textit{dotted}), and warm/hot ionized gas (\textit{dash-dotted}). The dust population of all galaxies is initially dominated by carbonaceous dust from SNe II stars, but this produces an extremely low D/Z. Carbonaceous dust from AGB stars eventually takes over as the dominant dust creation source as long as accretion remains inefficient due to low galactic metallicity, but D/Z changes little compared to the SNe II stardust-dominated population. If the galactic metallicity passes a critical threshold, accretion becomes efficient, becoming the dominant dust creation source and increasing the median D/Z in cool, dense gas. The onset of rapid growth via accretion for metallic iron, silicates, and carbonaceous dust species can be seen in the rapid increase of their corresponding species mass fraction. Overall, the growth of silicate dust via accretion produces the largest change in median D/Z.}
    \label{fig:subsample1_dust_evolution}
\end{figure*}

\begin{figure*}
    \plotsidesize{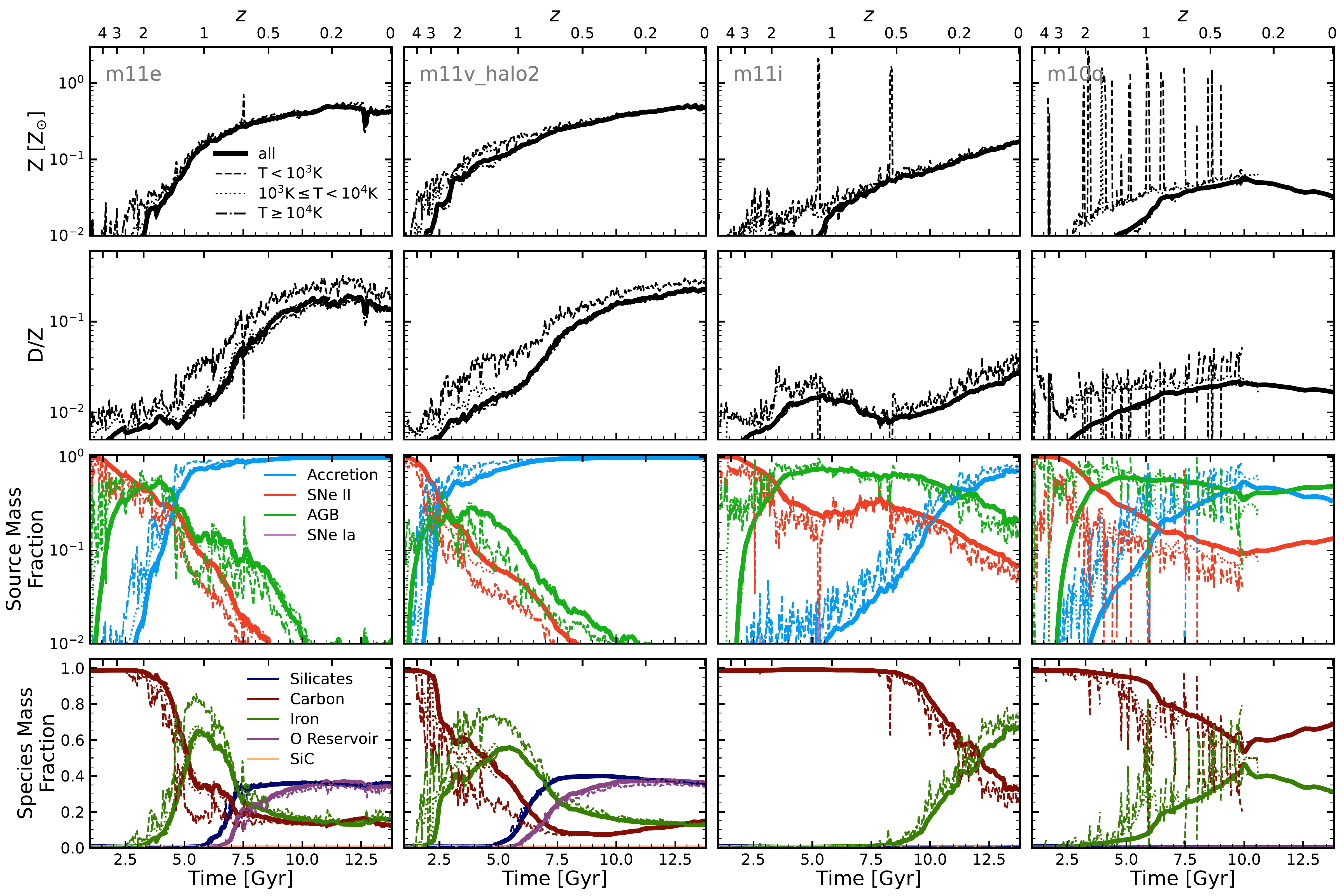}{0.99}
    \vspace{-0.25cm}
    \caption{Same as Fig.~\ref{fig:subsample1_dust_evolution} for {\bf m11e} 
    (\textit{left}), {\bf m11v\_halo2} (\textit{middle left}), and {\bf m11i} (\textit{middle right}) and {\bf m10q} (\textit{right}). Note that bursty star formation and resulting stellar feedback in the low-mass dwarf galaxies evacuates a majority of their cold gas which produces sudden increases in Z and decreases in dust for only cold gas.}
    \label{fig:subsample3_dust_evolution}
\end{figure*}

In Fig.~\ref{fig:subsample1_dust_evolution} \&~\ref{fig:subsample3_dust_evolution}, we show a detailed breakdown of each galaxy's metal and dust population evolution, including the median metallicity, median D/Z, dust creation source contribution, and dust species composition for all gas within 10 kpc of the galactic center. We also include the breakdown for each gas phase: cold neutral gas ($T<10^3$ K), warm neutral gas ($10^3$ K $\leq T < 10^4$ K), and warm/hot ionized gas ($T\geq10^4$ K). 
While the evolution of each galaxy varies considerably, their dust populations follow similar evolutionary trends, which we describe below.

{\bf 1. SNe-Dominated:} 
Initially, as the first massive stars die, SNe II are the dominant producer of dust mass. This `stardust' population is dominated by carbonaceous dust, arising from the high SNe II carbonaceous dust production efficiency adopted in our model  ($\delta^{\rm SNII}_{\rm carb}=15\%$). This produces an extremely low median D/Z $\sim0.01$. 

{\bf 2. AGB-Dominated:} 
As stellar populations age and low-mass stars transition to the AGB phase, dust production by AGB stars becomes a sizable component of the dust mass. However, this only dominates the dust population for the lowest-mass dwarf galaxies, which have little or extremely late dust growth via accretion.
For our model, low-metallicity AGB stars produce primarily carbonaceous dust and only x2-3 more dust than SNe II over the lifetime of a stellar population (see Fig.\ 2 in \citetalias{choban_2022:GalacticDustupModelling}). For these reasons, the median D/Z and dust composition exhibit little change between SNe and AGB-dominated dust regimes.

{\bf 3. Onset of Accretion:}
Once the gas within a galaxy reaches a `critical' metallicity threshold (i.e. key element number abundance), dust growth via gas-dust accretion becomes more efficient than dust destruction by SNe shocks. The exact `critical' metallicity varies between dust species (metallic iron, silicates, carbonaceous) due to differences in their key element abundances, physical properties, and life cycle processes, which we discuss in Sec.~\ref{sec:efficient_accretion}. Metallic iron is the first dust species to grow efficiently via accretion, then silicates, and last carbonaceous dust. Each successively increases the median D/Z, up to D/Z $\sim0.02$, $\sim0.2$, and $\sim0.3$ respectively.

{\bf 4. Build Up \& Equilibrium:} 
Once a dust species begins to grow efficiently via accretion, its dust mass builds up over time until an equilibrium is reached, which we discuss in Sec.~\ref{sec:dust_buildup}.
Once all dust species grow via accretion, the dust population of the galaxy reaches saturation, with the maximum amount of each dust species effectively forming. This produces an equilibrium dust population similar in composition to the Milky Way (${\rm D/Z}\sim0.4$ and Sil-to-C $\sim3{:}1$) and is relatively constant over time.

To gauge the accuracy of this predicted trend in dust population evolution, we must compare to observations of galactic dust population amount and chemical composition and observed dust population variability within galaxies.

\subsection{Element Depletions \& Aggregate D/Z} \label{sec:element_depletions}

Element depletions provide a detailed accounting for how much of each element is locked in dust and are a strong constraint for our model. Observationally, element depletions are measured using UV spectral absorption features from sight lines to bright background sources, usually O/B type stars or quasars. Sight line column densities for ionized refractory elements and neutral hydrogen are determined from fits to their absorption profiles, and relative abundances between each element and hydrogen are determined. These relative abundances are then compared to known reference abundances and any missing elements from the gas-phase are then assumed to be locked in dust. The gas-phase depletion of element X is represented logarithmically as 
\begin{equation} \label{eq:NH_depletion}
    \left[ \frac{\rm X}{\rm H} \right]_{\rm gas} = \log \left( \frac{N_{\rm X}}{N_{\rm H}} \right)_{\rm gas} - \log \left( \frac{N_{\rm X}}{N_{\rm H}} \right)_{\rm ref},
\end{equation}
and linearly as
\begin{equation}
    \delta_{\rm X} = 10^{\left[ {\rm X}/{\rm H} \right]_{\rm gas}},
\end{equation}
where $N_{\rm X}$ and $N_{\rm H}$ are the gas-phase column density of element X and column density of neutral hydrogen ($\NHn=N_{\rm \textsc{H\,i}}+2N_{\Hmol})$ respectively and $\left(N_{\rm X}/N_{\rm H} \right)_{\rm ref}$ is the assumed reference abundance. 

Due to the high spatial and spectral resolution needed for such observations, they are mainly limited to Milky Way, LMC, and SMC\footnote{Element depletions have been observed for damped Ly$\alpha$ systems (DLAs) via quasar absorption lines \citep{peroux_2020:CosmicBaryonMetal}. However, DLAs probe a variety of systems \citep{prochaska_1997:KinematicsDampedLymana,wolfe_2005:DampedLySystems,faucher-giguere_2011:SmallCoveringFactor,rhodin_2019:NatureStrongAbsorbers,stern_2021:NeutralCGMDamped} and their reference abundances can be difficult to determine due to dust depletions \citep{roman-duval_2022:METALMetalEvolution}. For these reasons, we forgo comparing to DLA depletion observations in this work.}.
Observations of Milky Way depletions compiled by \citet{jenkins_2009:UnifiedRepresentationGasPhase} show C, O, Mg, Si, and Fe depletion increases with increasing neutral H column density.
Mg, Si, and Fe depletion observations in the LMC \citep{roman-duval_2021:METALMetalEvolution} and SMC \citep{jenkins_2017:InterstellarGasphaseElement} show similar trends to those in the MW but are offset correlating with the relative differences in metallicity between the galaxies \citep{roman-duval_2022:METALMetalEvolutiona}. However, direct observations of C and O depletions are limited only to the Milky Way due to either saturated or extremely weak absorption lines, so C and O depletion trends for the SMC and LMC are usually inferred from the MW relation between Fe and C or O depletions respectively \citep{roman-duval_2022:METALMetalEvolutiona,roman-duval_2022:METALMetalEvolution}.

\begin{figure*}
    \plotsidesize{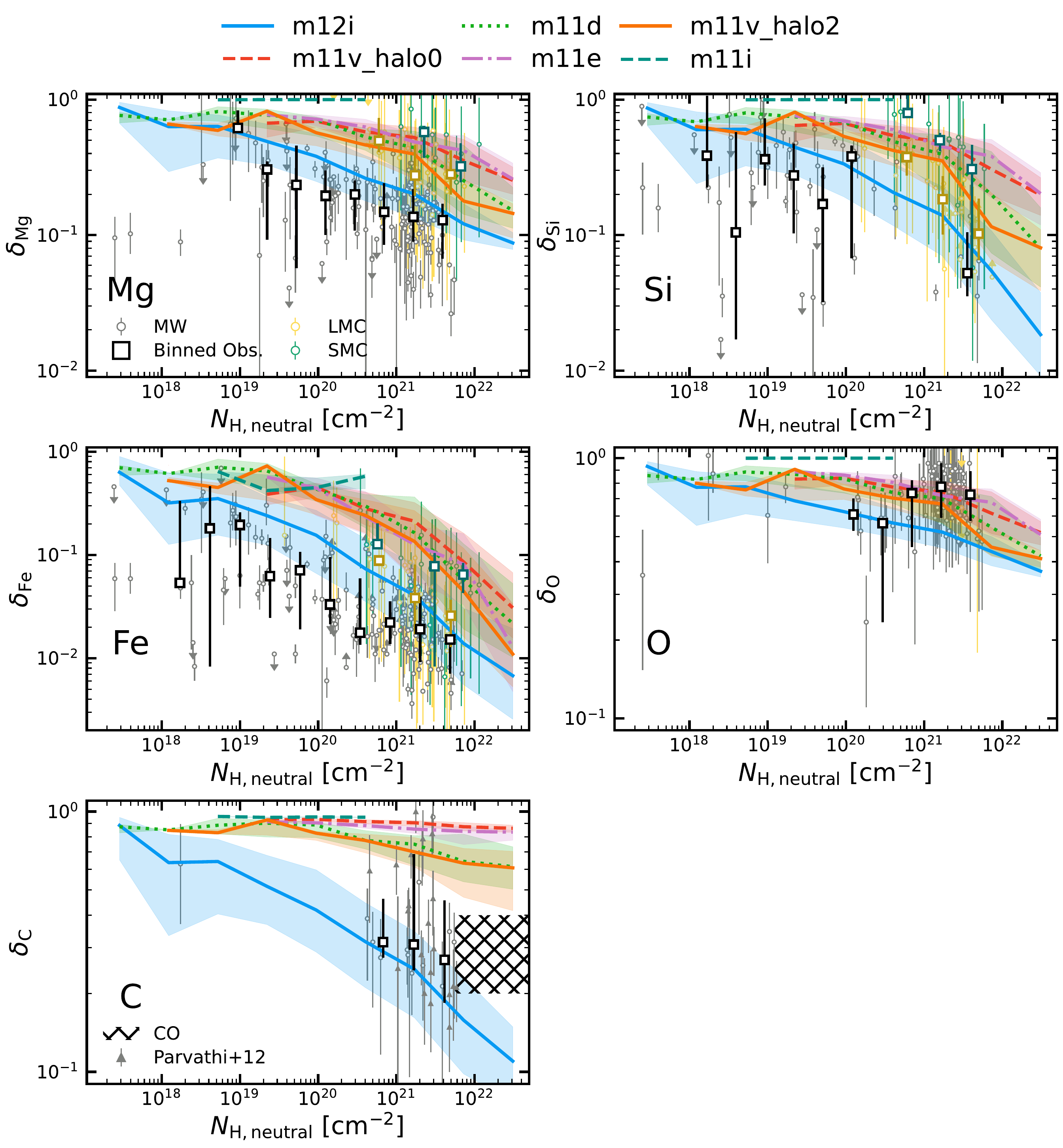}{0.99}
    \vspace{-0.25cm}
    \caption{Predicted median sight line C, O, Mg, Si, and Fe depletion versus $N_{\rm H,neutral}$ from in-disk sight lines for {\bf m12i} and face-on sight lines for {\bf m11v\_halo0}, {\bf m11d}, {\bf m11e}, {\bf m11v\_halo2}, and {\bf m11i} following the sight line methodology outlined in Appendix~\ref{Appendix_sightlines}. For each, 16-/84-percentile ranges are represented by shaded regions. We compare with observed sight line depletions within the Milky Way from \citet{jenkins_2009:UnifiedRepresentationGasPhase} ({\it grey circles}) and carbon depletions from \citet{parvathi_2012:ProbingRoleCarbon} ({\it triangles}), of the LMC from \citet{roman-duval_2022:METALMetalEvolutiona} ({\it yellow circles}), and of the SMC from \citet{jenkins_2017:InterstellarGasphaseElement} ({\it teal circles}). We include the binned median and 16-/84-percentile ranges for each data set ({\it squares}). 
    For C depletions, we also include a range of expected maximum depletions in dense environments ({\it hatched}) based on observations C in CO in the Milky Way.
    All galaxies (besides {\bf m11i}) produce a Mg, Si, Fe, and O relation which transitions from a shallow to steep slope for increasing $N_{\rm H,neutral}$, but each relation is offset roughly corresponding with each galaxy's median metallicity. The C relations exhibit a weaker transition and a larger offset, with the lowest metallicity galaxies ({\bf m11e}, {\bf m11v\_halo0}, and {\bf m11i}) producing a nearly flat relation. This is due to the `turn-off' of carbonaceous growth via accretion since these galaxies lie below the carbonaceous `critical' metallicity threshold. {\bf m11i} exhibits flat depletion relations for all but Fe since only metallic iron dust grows efficiently in this galaxy.
    Overall, the predicted trends are a good match with observations, but the lack of C or O depletion measurements outside the Milky Way paints an incomplete picture.}
    \label{fig:sight_line_element_depletion_NH}
\end{figure*}

To compare directly to these observations, we created a set of sight lines for each galaxy that originate from young star particles (formed within <10 Myr). For {\bf m12i}, the sight lines terminate within the galactic disk at a distance of $0.1-2$ kpc from the star to simulate depletion observations within the Milky Way. For the dwarf galaxies ({\bf m11v\_halo2}, {\bf m11d}, {\bf m11e}, {\bf m11v\_halo0}, and {\bf m11i}), the sight lines terminate $\sim$50 kpc outside the galaxy to simulate observations of the LMC and SMC. For each sight line we then calculated $\NHn$ and C, O, Mg, Si, and Fe depletions. 
The specifics of our sight line methodology are described in detail in Appendix~\ref{Appendix_sightlines}.
We binned these sight lines in logarithmic $\NHn$ bins and calculated the median values and 16-/84-percentiles for C, O, Mg, Si, and Fe depletions. 
The resulting relation between sight line element depletion and $\NHn$ for each element can be seen in Fig.~\ref{fig:sight_line_element_depletion_NH} for each galaxy.
We include the 243 observed Milky Way sight line depletions from \citet{jenkins_2009:UnifiedRepresentationGasPhase} and the carbon depletions from \citet{parvathi_2012:ProbingRoleCarbon}, the 32 sight line depletions of the LMC from \citet{roman-duval_2022:METALMetalEvolutiona}, and the 18 sight line depletions of the SMC from \citet{jenkins_2017:InterstellarGasphaseElement}.  We modify the carbon depletions from \citet{jenkins_2009:UnifiedRepresentationGasPhase}  and \citet{parvathi_2012:ProbingRoleCarbon} similar to Sec.\ 3.2.1 in \citetalias{choban_2022:GalacticDustupModelling} to correct for differences in depletion estimates determined from strong and weak \textsc{C\,ii} transition lines and total C abundances. We also include a range of expected maximum C depletions in dense environments based on observations of 20\%\ to 40\%\ of C in CO in the Milky Way \citep[e.g.][]{irvine_1987:ChemicalAbundancesMolecular, vandishoeck_1993:ChemicalEvolutionProtostellar, vandishoeck_1998:ChemicalEvolutionStarForming,lacy_1994:DetectionAbsorptionH2}.

\begin{figure*}
    \plotsidesize{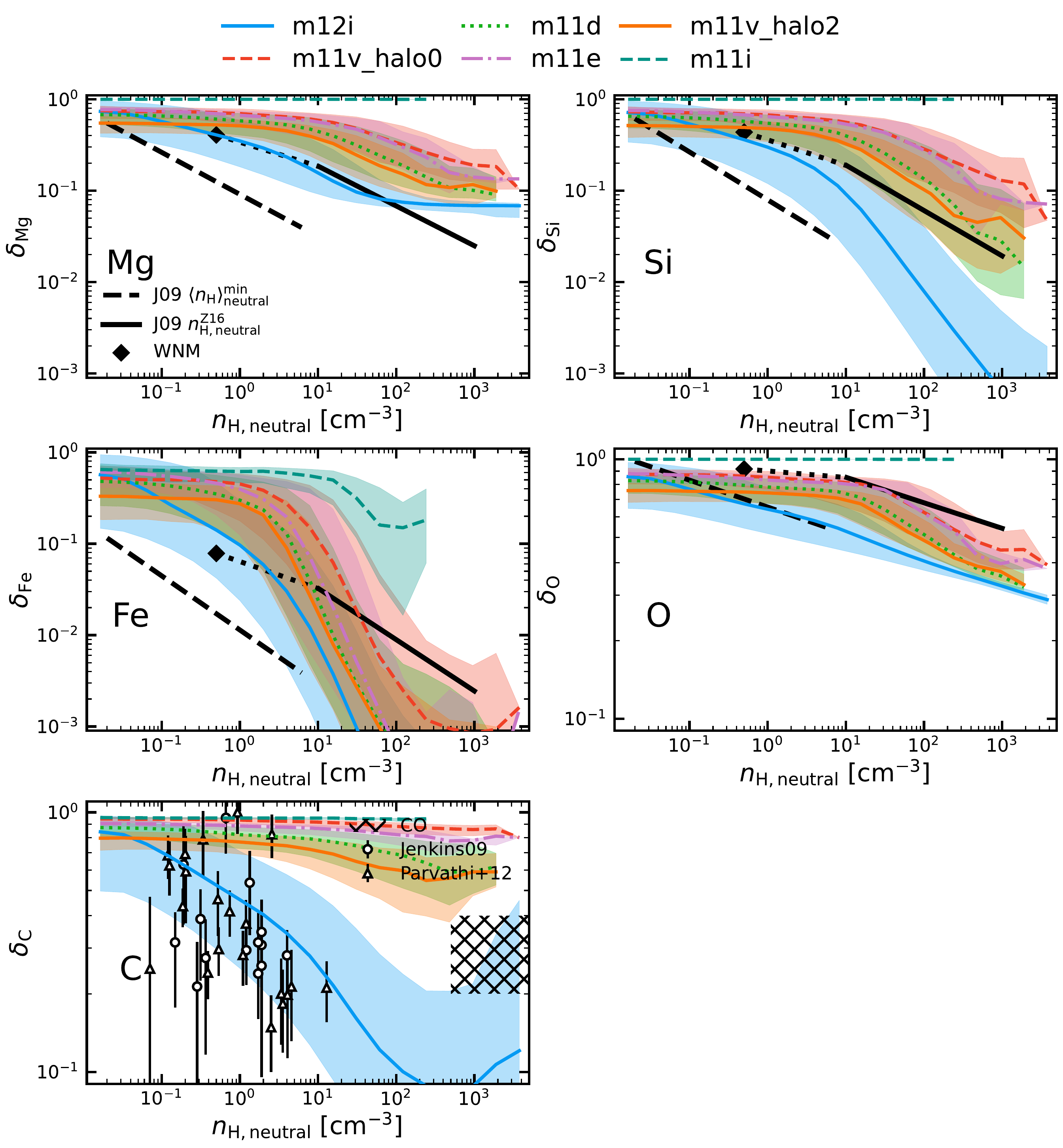}{0.9}
    \vspace{-0.25cm}
    \caption{Resulting C, O, Mg, Si, and Fe depletion versus $\nHn$ for gas within the 10 kpc of each galaxy at $z=0$. The 16-/84-percentile ranges are represented for each by shaded regions. 
    We compare with observed elemental depletion relations in the Milky Way from \citet{jenkins_2009:UnifiedRepresentationGasPhase} assuming mean sight line density is the physical density ({\it black-dashed}), which can be treated as a lower limit, and using \citet{zhukovska_2016:ModelingDustEvolution,zhukovska_2018:IronSilicateDust} mean sight line density to physical density fit ({\it black-solid}). For O, Mg, Si, and Fe we include estimates for the WNM depletions ({\it diamond}) along with an interpolation to the Jenkins' relation ({\it black-dotted}). 
    For C, we only include the individual sight line depletions from \citet{jenkins_2009:UnifiedRepresentationGasPhase} ({\it triangles}), decreased by a factor of 2, and \citet{parvathi_2012:ProbingRoleCarbon} ({\it circles}) as a lower bound. We also include a range of expected minimum C depletions in dense environments ({\it hatched}) based on observations of C in CO in the Milky Way.
    {\bf m12i}, {\bf m11v\_halo2}, {\bf m11d}, {\bf m11e} and {\bf m11v\_halo0} show similar offset trends for Mg, Si, O, and Fe which transition from a shallow to steep slope for increasing $n_{\rm H,neutral}$. Mg and C depletions show an additional transition to a flat slope at the highest densities. For Mg, this is due to the saturation of silicate dust, which has run out of available Si to grow further. For C, this is due to the rapid formation of CO, which takes up any remaining gas-phase C.
    {\bf m11i} only exhibits a slopped trend for Fe since only metallic iron grows efficiently.}
    \label{fig:element_depletion_nH}
\end{figure*}

We can also directly measure the depletion for each element as a function of physical gas density $\nHn$ since our simulations track the total abundance of each element locked in dust for all gas cells. This is valuable for understanding sight line depletions since individual sight lines probe various gas phases and sight lines with similar $\NHn$ can probe vastly different gas environments. We therefore bin the gas cells in logarithmic neutral gas density and calculate the median values and 16-/84-percentiles for C, O, Mg, Si, and Fe depletions. The resulting relation for each element depletion and neutral gas density, $\nHn$, is shown in Fig.~\ref{fig:element_depletion_nH} for each galaxy. 
We also include fits to the sight line depletion trends in the Milky Way from \citet{jenkins_2009:UnifiedRepresentationGasPhase} assuming mean sight line density is the physical density as a lower bound, and using \citet{zhukovska_2016:ModelingDustEvolution, zhukovska_2018:IronSilicateDust} mean sight line density to physical density fit (see \citetalias{choban_2022:GalacticDustupModelling} Sec.\ 3.2.2 for details).

O, C, Mg, Si, and Fe show similar depletion trends for all galaxies, transitioning from a shallow to steep slope with increasing density ($\NHn$ and $\nHn$) similar to observations. This is due to the cycling of gas into cold, dense regions, where metallic iron, silicates, and/or carbonaceous dust grow via gas-dust accretion, and out to hot, diffuse regions where all species are destroyed via sputtering and SNe shocks similar to the results found in \citetalias{choban_2022:GalacticDustupModelling} for an idealized MW.
The trends for C and Mg, in particular, flatten at high density ($\nHn > 10^2 \, \cmcubed$; not visible at high $\NHn$) for all galaxies but for different reasons. C depletions flatten due to our sub-resolution dense molecular cloud prescription. This prescription assumes gas-phase C is rapidly converted to CO in dense molecular gas, halting carbon dust growth. Mg depletions flatten due to the relative abundances of Si and Mg and silicate dust being the only depletion source of Mg in our model. Si is the key element for silicate dust for all galaxies, so there will always be some leftover gas-phase Mg once the maximum amount of silicate dust has formed\footnote{We note that the simulations presented in \citetalias{choban_2022:GalacticDustupModelling} predicted that Mg is the key element for silicates. This was due to the idealized nature of the simulations, whose initial conditions assumed uniform gas and stellar metallicities of $Z = Z_{\odot}$ and a uniform stellar age distribution over 13.8 Gyr. The metal abundances arising from such a simulation are invariably different from those produced by a realistic stellar population successively built over the age of the universe.}.

Comparing predicted trends between galaxies, all depletions exhibit a staggered offset from one another. These offsets roughly correlate with each galaxy's median gas-phase metallicity, with {\bf m11i} having the weakest and {\bf m12i} having the strongest depletion trends, similar to what is seen in the MW, LMC, and SMC. These offsets are caused by differences in the average gas-dust accretion rate for each galaxy, which scales with gas-phase metallicity (see Eq.20 in \citetalias{choban_2022:GalacticDustupModelling}). Galaxies with higher average gas-phase metallicity will have higher gas-dust accretion rates in dense environments and, therefore, stronger element depletions in these environments. C depletion trends exhibit the largest offset between galaxies and decreasing slopes at high densities. This decrease is so dramatic that only {\bf m12i}, {\bf m11d}, and {\bf m11v\_halo2} produce noticeable increases in depletion with increasing density while {\bf m11e}, {\bf m11v\_halo2}, and {\bf m11i} have entirely flat relations.
This is due to the turn-off of efficient carbonaceous dust growth via accretion. 
Galaxies with metallicities below a `critical' metallicity threshold needed for efficient carbonaceous dust growth will produce flat depletion relations set by the creation of dust in SNe and AGB winds. This can also be seen in the flat O, Mg, and Si depletion relations for {\bf m11i} due to the lack of silicate dust growth via accretion.

\begin{figure*}
    \plotsidesize{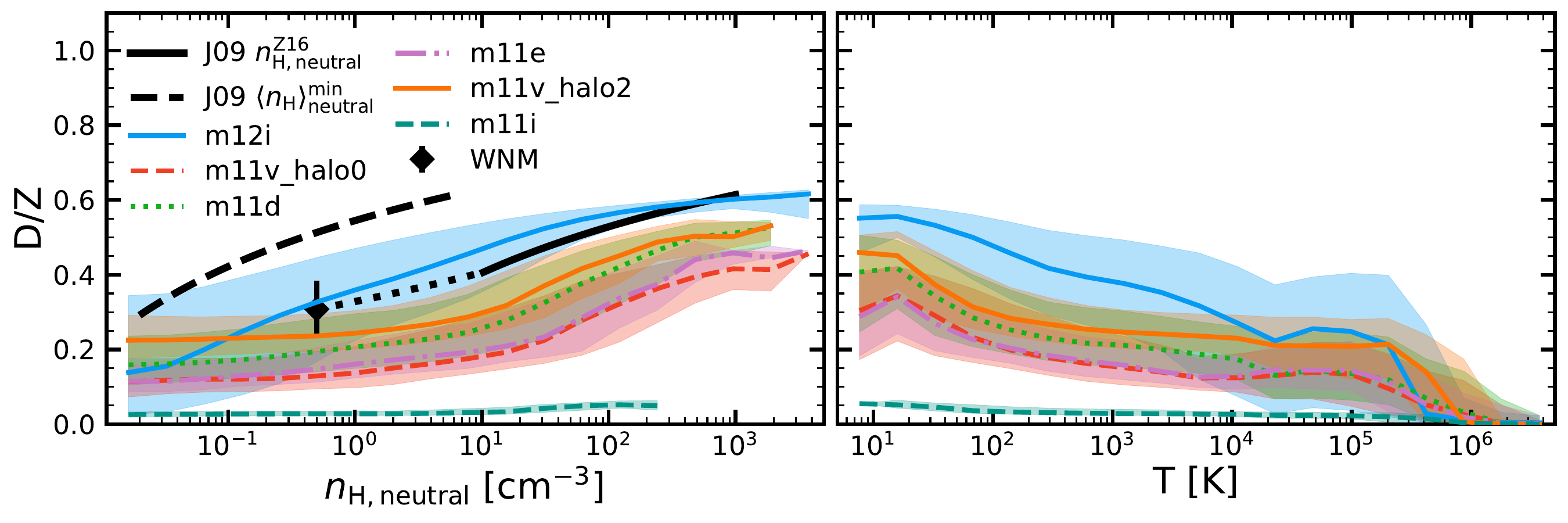}{0.99}
    \vspace{-0.25cm}
    \caption{Resulting D/Z versus $\nHn$ ({\it left}) and temperature ({\it right}) for gas within 10 kpc for each galaxies at $z=0$. All galaxies with efficient dust growth via accretion produce a sloped relation which increases with increasing $\nHn$ and decreasing temperature. All galaxies with efficient silicate growth via accretion also have sizable D/Z even in low density, hot gas ($T>10^4$ K or $\nHn<1 \, \cmcubed$), which is exposed to SNe shocks.
    For gas with $T>10^5$K, D/Z quickly drops due to the onset of efficient dust destruction via thermal sputtering.}
    \label{fig:DZ_vs_nH_T}
\end{figure*}

We further aggregate the element depletions into total D/Z to determine the resulting distribution of D/Z with $\nHn$ and temperature for each galaxy. We binned the gas cells in logarithmic neutral gas density and temperature bins, calculating the median D/Z and 16-/84-percentiles shown in Fig.~\ref{fig:DZ_vs_nH_T}. Aggregating the observed Milky Way element depletions, we include an upper bound on the expected D/Z relation with $\nHn$ along with an estimate for the WNM\footnote{To account for the uncertainty in the observed WNM D/Z owing to the lack of measured sight line C depletions in this regime we include error bars representing 80\% (assuming 20\% in CO) of C or no C in dust.} 
based on depletions from \cite{jenkins_2009:UnifiedRepresentationGasPhase}. We also include a reasonable estimate of D/Z in dense environments following the relation from \cite{zhukovska_2016:ModelingDustEvolution,zhukovska_2018:IronSilicateDust}.

The exhibited D/Z relations with $\nHn$ mirror the individual element depletion trends for each galaxy, transitioning from a shallow to steep slope with increasing density and offset from one another in proportion to each galaxy's relative metallicity. This transition is less prominent in {\bf m12i}, but it matches reasonably well with observations of the Milky Way. In regards to temperature, the resulting D/Z relation for all galaxies can be broken into three regimes. {\bf (i) Cool/Warm Gas ($\bf{T<10^4}$K):} As hot gas cools, gas-dust accretion eventually `turns on' and dust begins to grow, steadily increasing D/Z with decreasing temperature. Our model assumes this `turn on' point occurs when $T<300$K, which can be seen as a moderate increase in D/Z.
{\bf (ii) Ionized Gas ($\bf{T=10^4-10^5}$ K):} This regime is primarily comprised of gas that has recently experienced SNe feedback, with dust residing in gas cells immediately around the SNe being destroyed. This produces a relatively flat D/Z for all galaxies, which lies at $\sim$40\% of the maximum D/Z at low temperatures for each galaxy. This matches surprisingly well with the set dust destruction efficiency of $\bar{\epsilon}\approx40\%$ assumed for gas shocked to $v_{s}\geq100$ km/s in our SNe dust destruction routine. This indicates that, on average, our routine predicts all gas in this regime has been shocked to $v_{s}\geq100$ km/s once.
However, the large scatter in D/Z indicates this varies considerably between gas cells, with some being shocked multiple times and others not shocked at all.
{\bf (iii) Coronal Gas ($\bf{T>10^5}$ K):} Dust destruction via thermal sputtering becomes efficient in this regime, with D/Z rapidly dropping with increasing temperature. For gas with $T\gtrsim10^6$ K, on average, all dust is destroyed.

\subsection{Extragalactic Dust Emission} \label{sec:dust_emission}

Direct measurement of element depletions in galaxies beyond the Milky Way and its satellites is currently not possible, aside from DLAs. An alternative method to probe galactic dust populations is through the combination of multi-wavelength estimates of dust mass, gas mass, and metallicity. 
This method infers a total dust mass by fitting dust emission models to observed dust emission spectra.
The assumed dust model can vary in complexity from a simple modified black body to physical dust models with specific dust grain sizes and chemical compositions, affecting the total dust mass inferred \citep{chastenet_2021:BenchmarkingDustEmission}. 
This method was applied to a large sample of galaxies by \citet{remy-ruyer_2014:GastodustMassRatios}, who studied the galaxy-integrated D/Z for 126 local galaxies, including dwarf, spiral, and irregular type galaxies, covering a large metallicity range ($\OH=7.1-9.1$; $Z\approx0.02-2.3\,\Zsol$). 
\citet{devis_2019:SystematicMetallicityStudy} further expanded upon this sample, adding $\sim500$ local DustPedia galaxies, which cover a more limited metallicity range ($\OH=7.9-8.7$; $Z\approx0.15-0.9\,\Zsol$) and are mainly limited to spiral galaxies.
\citet{chiang_2021:ObservationsSpatiallyResolved,chiang_2021:ResolvingDusttoMetalsRatio,chiang_2023:KpcscalePropertiesDust} extended this method to ${\sim}2$ kpc spatially-resolved studies of individual galaxies, compiling spatially resolved relations between D/Z and local environments for 46 nearby galaxies.
Their sample is limited to primarily spiral galaxies with $M_{\rm star}>10^9 \Msol$ and $12+{\rm log_{10}(O/H)} \gtrsim 8.4$ ($Z \gtrsim 0.5\,\Zsol$).  
Recently, \citet{clark_2023:QuestMissingDust} utilized this method with improved Herschel data
to produce high resolution (14 - 144 pc) maps of dust-to-neutral-gas (D/H$_{\rm neutral}$) ratios for Local Group galaxies (LMC, SMC, M33, and M31).

\subsubsection{Galaxy-Integrated D/Z}

\begin{figure*}
    \plotsidesize{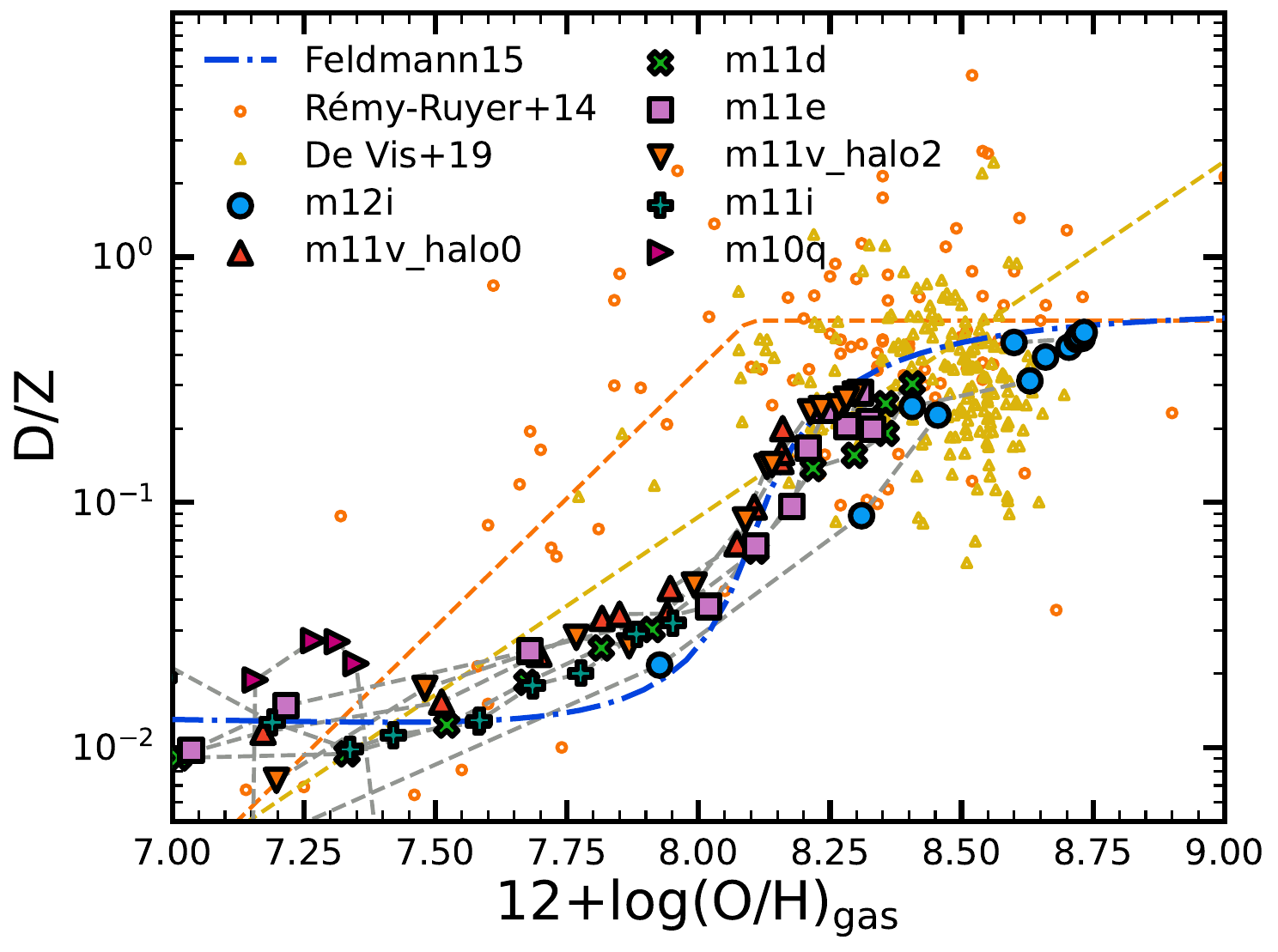}{0.75}
    \vspace{-0.25cm}
    \caption{Resulting relation between galaxy integrated D/Z and metallicity for our galaxies over time. Each connected point is +1 Gyr apart in time, starting at $\sim$3 Gyr up to present day. The galaxy-integrated D/Z is set as the median D/Z for cool, neutral ($T{<1}000$ K) gas, similar to the type of environments probed by observations of dust SEDs. The metallicity is set as the median gas-phase oxygen abundance (only O not depleted into dust) for gas with $7000<T<15000$ and $\nH>0.5$ cm$^{-3}$, similar to regions probed by nebular emission lines used in galactic observations. $\OH$ is also offset by -0.2 to account for the differences in assumed oxygen abundances between \citet{anders_1989:AbundancesElementsMeteoritic} and \citet{asplund_2009:ChemicalCompositionSun}. We compare with galaxy-integrated observations of local galaxies from \citet{remy-ruyer_2014:GastodustMassRatios} ({\it orange circles}) and \citet{devis_2019:SystematicMetallicityStudy} ({\it gold triangles}), along with the fits presented in both studies ({\it dashed lines}).
    We also include predictions from the equilibrium chemical and dust model presented in \citet{feldmann_2015:EquilibriumViewDust} ({\it blue dash-dotted}).
    All galaxies start with low D/Z and Z, being dominated by stardust production. For galaxies that surpass $\OH\sim7.9$, D/Z rises steeply due to the onset of efficient silicate, and later carbonaceous, dust growth via accretion. The differences in the steepness of this relation are due to the time it takes for the dust population to build up through successive cycling of gas into and out of cold, dense clouds.
    {\bf m12i} also experiences a temporary decrease in the metallicity due to a minor merger which decreases the median metallicity of warm gas as seen in Fig.~\ref{fig:subsample1_dust_evolution}.}
    \label{fig:galaxy_int_DZ}
\end{figure*}

We first show the resulting evolution between galaxy-integrated D/Z and metallicity for each simulated galaxy in Fig.~\ref{fig:galaxy_int_DZ}, plotting D/Z and $\OH$ in 1 Gyr intervals from $\sim$3 Gyr to present day. We compare against the galaxy-integrated observations from \citet{remy-ruyer_2014:GastodustMassRatios} and \citet{devis_2019:SystematicMetallicityStudy}\footnote{The offset between these observational studies is, in part, due to \citet{devis_2019:SystematicMetallicityStudy} unique definition of ${\rm D/Z}=\Sigma_{\rm dust}/(\Sigma_{\rm dust}+\Sigma_{\rm metals})$ to account for possible depletion of metals into dust. } 
along with their fitted relations. 
We define the galaxy-integrated D/Z as the median D/Z for cool, neutral ($T<1000$ K) gas, following the assumption that galaxy-integrated dust emission studies primarily probe dense environments.
In particular, \citet{aniano_2012:ModelingDustStarlight,aniano_2020:ModelingDustStarlight} directly compared observations of spatially-resolved and galaxy-integrated dust emission SEDs for individual galaxies and found that a majority of galaxy-integrated dust emission is produced by dust in dense regions exposed to diffuse radiation fields.
\citet{chiang_2021:ResolvingDusttoMetalsRatio,chiang_2023:KpcscalePropertiesDust} also found that regions with good SNR (high dust emission) correlate with regions of CO detection (molecular gas), which further supports this assumption. 
We define the galaxy-integrated metallicity as the median $\OH$ for gas with $7000<T<15000$ and $\nH>0.5 \,\cmcubed$ to match the properties of nebular regions typically probed by empirical strong emission line methods used in these studies \citep[e.g.][]{pilyugin_2016:NewCalibrationsAbundance}.
We also account for the depletion of O into dust by only considering gas-phase O instead of total (gas+dust) O abundance and include a -0.2 offset to correct for differences in reference O abundances assumed in our simulations \citep{anders_1989:AbundancesElementsMeteoritic} and observations \citep{asplund_2009:ChemicalCompositionSun}. We investigate other definitions of $\OH$ in Appendix~\ref{Appendix_OH} and find that accounting for only gas-phase O has the largest effect, but this only produces an offset of $\lesssim 0.2$ primarily for high values of $\OH$. However, this depends entirely on our model's prescription for O depletion into dust.
We also include the predicted relation between D/Z and metallicity from the equilibrium chemical and dust analytical model presented in \citet{feldmann_2015:EquilibriumViewDust}, which tracks the evolution of a single, chemically-ambiguous dust species and accounts for stardust production, dust growth in the ISM, dust destruction by SNe shocks, and dust dilution by inflowing, pristine gas. 
We specifically plot the results from their model with slight adjustments to their fiducial parameters. We set the maximum depletion limit for all metals to $f^{\rm dep}=0.6$ to match the maximum D/Z predicted by our model. We modify the dust injection via AGB stars and SNe II of a
stellar population to $y_{\rm D}=2\times10^{-3}$ to match our typical D/Z for stardust dominated galaxies. Finally, we set the ratio between the molecular gas depletion by star formation timescale and the dust growth via accretion timescale to $\gamma = 1.3\times 10^4$ to match our model's predicted rapid increase in D/Z above a critical metallicity threshold.

At low metallicity, all of our galaxies exhibit similar D/Z relations. 
These galaxies are dominated by stardust production, which yields very low D/Z $\sim10^{-2}$ that is roughly constant with metallicity. Metallic iron dust growth becomes efficient for galaxies above $\OH\sim7.6$, increasing D/Z by roughly a factor of two.
Above $\OH>7.9$, the relations begin to diverge due to the onset of efficient silicate and later carbonaceous dust growth. Some galaxies ({\bf m11v\_halo2} and {\bf m11v\_halo0}) follow the predicted equilibrium relation from \citet{feldmann_2015:EquilibriumViewDust} to surprising accuracy. Others ({\bf m12i}, {\bf m11d}, and {\bf m11e}) fall below this relation to varying degrees and at different metallicities, producing a maximum scatter in D/Z of ${\sim}0.5$ dex at $\OH\sim8.3$. These varying relations are caused by delays in the buildup of dust mass once gas-dust accretion becomes efficient, which we discuss in Sec.~\ref{sec:dust_buildup}.
At high metallicity, the relations begin to reconverge onto the equilibrium track where D/Z saturates at $\sim$0.4. At these metallicities, dust growth is so strong that a majority of all refractory elements have been locked into dust. Another interesting feature is the apparent `backslides' in metallicity shown by {\bf m12i}. These correspond to flyby and merger events which cause the infall of low-metallicity gas. This infalling gas decreases the metallicity of warm/hot gas, as can be seen in Fig.~\ref{fig:subsample1_dust_evolution}, including the nebular gas we use for our $\OH$ definition.
Even with our small sample of galaxies, we are able to recreate a reasonable amount of the scatter in D/Z for certain metallicities, which could explain some of the observed scatter\footnote{This scatter is partly due to observational effects, such as how a single galactic metallicity is assigned for galaxies which have radial metallicity gradients, which can result in unphysical D/Z $>1$.}. However, compared to the observed sample from \citet{remy-ruyer_2014:GastodustMassRatios}, our relations appear to be shifted to higher metallicity, indicating our `critical' metallicities may be too high.

Compared to other implementations, our findings are in good agreement with the majority of cosmological simulations \citep{aoyama_2018:CosmologicalSimulationDust,hou_2019:DustScalingRelations,li_2019:DusttogasDusttometalRatio, parente_2022:DustEvolutionMUPPI} and semi-analytical models \citep{popping_2017:DustContentGalaxies,vijayan_2019:DetailedDustModelling}. These all predict a similar s-shape trend between D/Z and metallicity due to the transition from stardust-dominated to accretion-dominated dust populations with increasing metallicity. We also produce a comparable amount of scatter in D/Z despite our limited sample size.
However, our simulations predict a slightly more gradual rise in D/Z due to differing `critical' metallicities for each dust species.

\subsubsection{Spatially-Resolved D/Z and D/H}

\begin{figure*}
    \plotsidesize{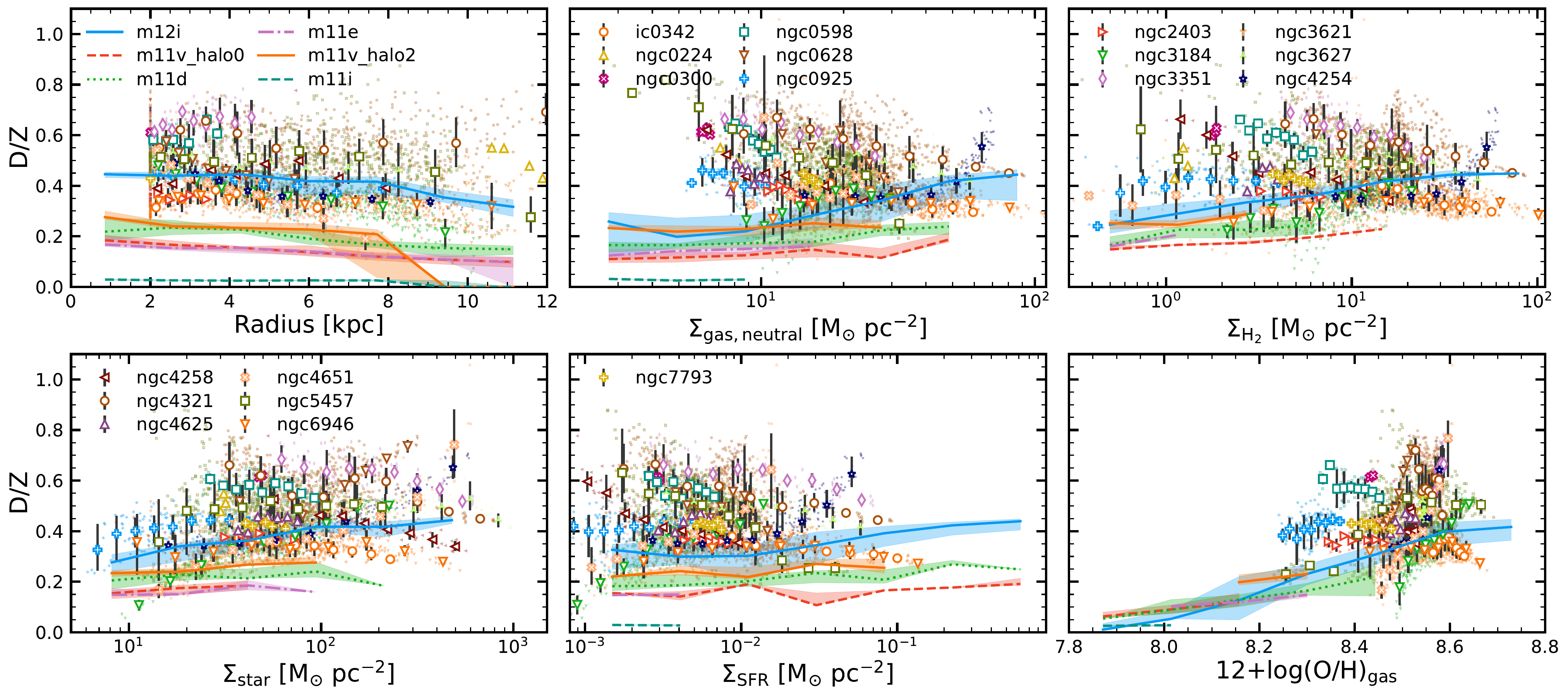}{0.99}
    \vspace{-0.25cm}
    \caption{Relation between median D/Z ratio and galactocentric radius ({\it top left}), neutral gas surface density ({\it top middle}), molecular gas surface density ({\it top right}), stellar surface density ({\it bottom left}), star formation rate surface density ({\it bottom middle}), and gas-phase oxygen abundance ({\it bottom right}) in 2 kpc face-on bins for our galaxies at $z=0$. We compare with dust emission-based observations of spatially-resolved D/Z of local galaxies from \citet{chiang_2023:KpcscalePropertiesDust}, which is an extension of the technique used in \citet{chiang_2021:ResolvingDusttoMetalsRatio}, with 2 kpc resolution and $\alpha_{\rm CO}^{\rm B13}$ conversion factor. All of our simulated galaxies produce a weakly slopped D/Z relation, which decreases with galactocentric radius and increases with density and metallicity. {\bf m12i} falls near the middle of, and the dwarf galaxies fall at the bottom or below the observed range for all galactic properties, but this is not unexpected since the observed sample includes only one dwarf galaxy. Our results disagree with the observed relations for $\Sigma_{\rm gas, neutral}$ and $\Sigma_{\Hmol}$, which suggest D/Z decreases in denser environments. However, these observations disagree with high-resolution observations \citep{clark_2023:QuestMissingDust}, so this may be due to resolution effects.}
    \label{fig:gal_resolved_DZ}
\end{figure*}

We further compare our simulations at $z=0$ to the spatially resolved observations of \citet{chiang_2023:KpcscalePropertiesDust}\footnote{Note that the specific $\alpha_{\rm CO}^{\rm B13}$ data products shown here are not explicitly presented in \citet{chiang_2023:KpcscalePropertiesDust}, but can be found in \citet{chiang_2021:ObservationsSpatiallyResolved}.} in Fig.~\ref{fig:gal_resolved_DZ}, examining the relation between D/Z and inclination corrected galactocentric radius, neutral gas surface density $(\Sigma_{\rm gas,neutral})$, molecular gas surface density $(\Sigma_{\Hmol})$, stellar surface density $(\Sigma_{\rm star})$, star formation rate surface density $(\Sigma_{\rm SFR})$, and $\OH_{\rm gas}$\footnote{Our spatially-resolved O abundances follow the same definition as defined for galaxy-integrated O abundances.}. 
We limit our comparison to the 17 galaxies in their 2 kpc resolution sample that have direct metallicity measurements and the derived D/Z values using the \citet{bolatto_2013:COtoH2ConversionFactor} $\alpha_{\rm CO}$ prescription ($\alpha_{\rm CO}^{\rm B13}$), which \citet{chiang_2021:ResolvingDusttoMetalsRatio} argued yields the most reasonable D/Z.
To match the observational resolution, we bin each simulation in 2 kpc face-on square pixels and calculate D/Z $=\Sigma_{\rm dust}/\Sigma_{\rm metals}$ for each pixel. We then bin these pixels across each property and calculate the median D/Z values and 16-/84 percentiles for each\footnote{While the observations from \citet{chiang_2023:KpcscalePropertiesDust} primarily probe $\Hmol$-dominated regions (employ a minimum $I_{\rm CO}$ detection threshold), we do not include a molecular gas mass fraction ($f_{\Hmol}$) cutoff for simulated pixels. We find that including such a cutoff has little effect on the resulting relations for {\bf m12i}, primarily truncating the relations at lower surface densities and $\OH$, but excludes the majority of pixels for our dwarf galaxies.}.

All galaxies produce a weakly slopped relation between D/Z and all galactic properties. Specifically, increasing D/Z with decreasing galactocentric radius and increasing density and metallicity. All relations are roughly offset downwards by each galaxy's relative metallicity, besides the relation between D/Z and $\OH$, which suggest all galaxies follow the same trend across their respective metallicity ranges.
Compared to observations, {\bf m12i} falls near the middle of the observed range for all galactic properties. The dwarf galaxies fall either at the bottom or below the range of observations, but the observational sample includes only one dwarf galaxy (M33/NGC598), so this is not unexpected.
While our simulations disagree with the observed trends of decreasing D/Z with increasing $\Sigma_{\rm gas,neutral}$ and $\Sigma_{\Hmol}$ for many galaxies, this decreasing trend is likely due to observational biases at low column densities which can increase the expected D/Z. Specifically, CO-dark $\Hmol$ gas is not accounted for, which will decrease the expected $\Sigma_{\rm gas,neutral}$/$\Sigma_{\Hmol}$ and thus increase D/Z. Furthermore, SNR pixel cuts for dust emission will preferentially exclude low D/Z pixels for low $\Sigma_{\rm gas,neutral}$/$\Sigma_{\Hmol}$. We also highlight that high-resolution dust emission observations from \citet{clark_2023:QuestMissingDust} exhibit the opposite trend, with D/Z increasing with $\Sigma_{\rm gas,neutral}$.

\begin{figure}
    \centering
    \includegraphics[width=0.99\columnwidth]{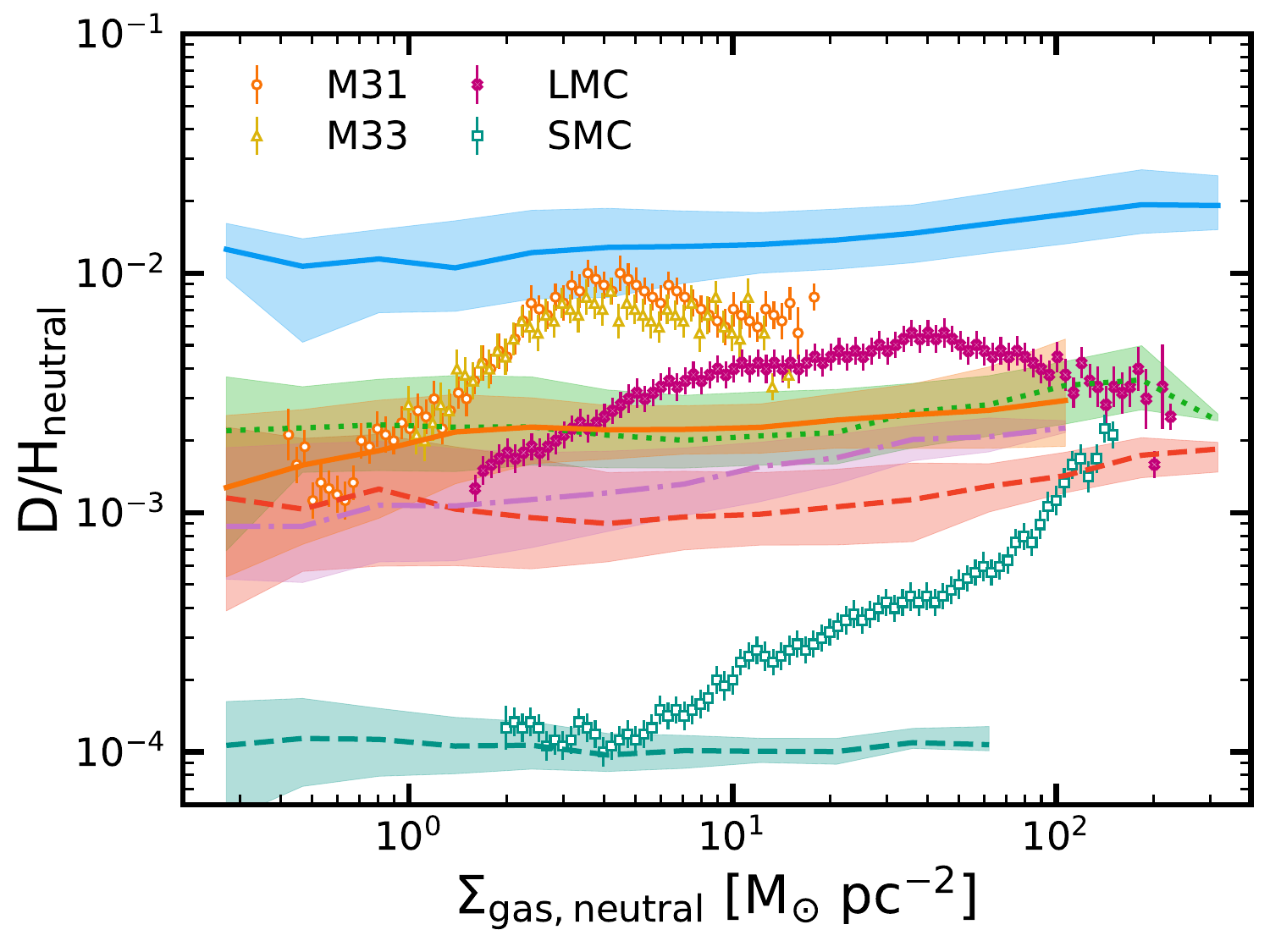}
    \vspace{-0.25cm}
    \caption{Resulting D/H$_{\rm neutral}$ versus $\Sigma_{\rm gas, neutral}$ for 144 pc resolution pixels from face-on projections of each galaxy at $z=0$. Only pixels with neutral gas mass fraction ${>}50\%$ are included since pixels dominated by ionized gas produce artificially high D/H$_{\rm neutral}$ at low $\Sigma_{\rm gas, neutral}$. We compare to deprojected `face-on' observations of M33, M31, LMC, and SMC from \citet{clark_2023:QuestMissingDust}, which vary in resolution from 14-144 pc. Note the error bars presented for these observations are the uncertainty of the median and not the scatter. The average standard deviation for D/H$_{\rm neutral}$ across all galaxies is 0.8 dex, but well-behaved and Gaussian. We caution that these observations may underpredict D/H$_{\rm neutral}$, when compared to element depletion observations. Overall, our simulations predict D/H$_{\rm neutral}$ increases with surface density and match the typical values observed at high $\Sigma_{\rm gas, neutral}$. However, they underpredict the steepness of this relation, suggesting our SNe dust destruction prescription may be too weak.}
    \label{fig:DH_vs_SigmaHn}
\end{figure}

We similarly compare our simulations to high-resolution observations of dust-to-neutral-gas mass ratio (D/H$_{\rm neutral}=\Sigma_{\rm dust}/\Sigma_{\textsc{Hi}+\Hmol}$) for Local Group galaxies from \citet{clark_2023:QuestMissingDust} in Fig.~\ref{fig:DH_vs_SigmaHn}. We utilize the same face-on projection technique outlined above but with 144 pc pixel sizes to match the lowest observational resolution in their sample.
We also only consider pixels with a neutral gas mass fraction $f_{\rm neutral}>0.5$ to avoid \textsc{Hii}-dominated regions since \citet{clark_2023:QuestMissingDust} does not include \textsc{Hii} gas in their gas mass accounting.
Overall, our simulations predict D/H$_{\rm neutral}$ increases with surface density for galaxies in which silicate and/or carbonaceous gas-dust accretion is efficient and match the typical D/H$_{\rm neutral}$ observed at high $\Sigma_{\rm gas,neutral}$ for all observed galaxies\footnote{The apparent downturn in observed D/H$_{\rm neutral}$ at the highest surface densities for the LMC, M31, and M33 is most likely caused by observational biases and model assumptions, and not a physical decrease in dust in these environments. We point the reader to Appendix B in \citet{clark_2023:QuestMissingDust} for a detailed breakdown of possible culprits.}.
However, the steepness of this relation is underpredicted, with  D/H$_{\rm neutral}$ only increasing by a factor of ${\lesssim}2$ in our simulations compared to the ${>}3$ factor observed for all galaxies. This suggests that our model's SNe dust destruction prescription is too weak, leaving too much dust in diffuse gas. However, we cannot entirely rule out our model predictions since the observations have an average standard deviation of 0.8 dex across all galaxies.
We also note that when compared to the expected D/H$_{\rm neutral}$ trends derived from element depletions (again using inferred C and O depletions), these observations underpredict D/H$_{\rm neutral}$ for the SMC but agree with the LMC.
Our simulations also indicate that efficient silicate dust growth may have occurred relatively recently in the SMC, given its extremely steep relation that overlaps with  {\bf m11v\_halo0} at high $\Sigma_{\rm gas, neutral}$ and {\bf m11i} at low $\Sigma_{\rm gas, neutral}$.

We caution that our approximation of observables begins to break down at these high resolutions. Specifically, individual phases of the ISM are resolved, resulting in pixels that primarily probe diffuse, ionized gas. These pixels have misleadingly high D/H$_{\rm neutral}$ at low neutral surface densities since a majority of dust resides in ionized gas. This produces an upswing in D/H$_{\rm neutral}$ at the lower end of our predicted relation and is even more pronounced when we do not include the $f_{\rm neutral}>0.5$ cutoff.
Whether observations would detect these pixels due to their possibly weak IR emissions is unknown and would require the creation of mock dust SEDs using radiative transfer codes. Due to this complexity, we save a more detailed comparison to spatially-resolved dust emission studies for future work.


\section{Discussion} \label{sec:discussion}

For all but the most metal-poor galaxies in our simulation suite, gas-dust accretion is the dominant producer of dust mass for a majority of each galaxy's life. Therefore, understanding when gas-dust accretion becomes efficient within a galaxy and when that galaxy's dust population transitions from a low to high D/Z ratio are critical for investigating galactic dust evolution.
However, these two transitions have typically been lumped together under the moniker of `critical' metallicity \citep[e.g.][]{zhukovska_2014:DustOriginLatetype,feldmann_2015:EquilibriumViewDust}. 
Below we break down these two transitions in our simulations, redefining the meaning of `critical' metallicity, and investigate their implications for observations.

\subsection{Efficient Gas-Dust Accretion} \label{sec:efficient_accretion}

The transition to efficient gas-dust accretion for a dust species occurs when, on average, more dust mass is created by accretion in dense regions than is destroyed by destruction processes (SNe, thermal sputtering, astration) or removed from the galaxy by outflows. 
While this implies that numerous, interwoven processes on both a local and galactic scale determine this transition, in practice, the galaxy-averaged metallicity is found to be the primary determinator. In particular, a `critical' metallicity threshold has been proposed above which the predicted galactic D/Z rapidly increases due to accretion \citep[e.g.][]{zhukovska_2014:DustOriginLatetype,feldmann_2015:EquilibriumViewDust,triani_2020:OriginDustGalaxies}. For our purposes, we define the `critical' metallicity as the point at which the species mass fraction of a given dust species begins to increase as seen in Fig.~\ref{fig:subsample1_dust_evolution} \&~\ref{fig:subsample3_dust_evolution}.
Using this definition, all of our simulated galaxies show the same `critical' metallicity threshold, above which the mass of a given dust species begins to increase. They all also indicate that each dust species has its own `critical' metallicity, with $\Zcrit\sim0.05\Zsol$ for metallic iron, $\Zcrit\sim0.2\Zsol$ for silicates, and $\Zcrit\sim0.5\Zsol$ for carbonaceous dust.
Three factors cause these differences in `critical' metallicity:

{\bf (1) Key Elements}: For a dust species with a set chemical composition, its accretion rate scales with the number abundance of that species' key element. For metallic iron this is Fe, for silicates this is Si\footnote{Depending on the relative abundances of Mg and Si, Mg can be the key element for silicates. For our simulations, we find Si is the key element for nearly the entire history of each galaxy.}, and for carbonaceous this is C. 
Assuming solar abundances (\citealt{anders_1989:AbundancesElementsMeteoritic} or \citealt{asplund_2009:ChemicalCompositionSun}), C is 10 times more abundant than either Si or Fe. 
Atomic C is also lighter, so it will have thermal velocities 1.5 and 2.2 times greater than Si and Fe respectively.
If we assume dust grows purely from hard-sphere type encounters, and all dust species are the same in all other respects, then carbonaceous dust should grow $\sim$15 times faster than silicates and  $\sim$22 times faster than metallic iron dust, which is similar to the predictions of \citet{granato_2021:DustEvolutionZoomin}.

{\bf (2) Physical Properties:} Differences in physical properties between dust species (i.e. grain sizes, geometry, grain charging, sticking efficiency) can alter their effective accretion rates. In particular, our model assumes different average grain sizes ($\left<a\right>_{3}$; see Eq. 3 in \citetalias{choban_2022:GalacticDustupModelling}) for each dust species due to differences in grain size distributions and effects of Coulomb enhancement\footnote{Coulomb enhancement arises from a grain size dependent electrostatic enhancement factor which accounts for the change in interaction cross section between ionized gas-phase metals and charged dust grains.} in atomic and diffuse molecular gas. 
Overall this produces a relative difference in accretion timescales of 1:150:900 in atomic/diffuse molecular gas for metallic iron, silicates, and carbonaceous dust, respectively, and 1:10:10 in dense molecular environments.

{\bf (3) Life Cycles:} Differences in dust life cycle processes (creation, growth, destruction) between species affect the net change in dust mass during one cycle of a gas parcel into and out of dense regions. For carbonaceous dust, our model accounts for the rapid formation of gas-phase CO in dense molecular environments, which halts dust growth. This decreases the amount of time carbonaceous dust has to grow compared to other dust species. For metallic iron, we assume a fraction of the dust population is locked inside silicate dust as inclusions ($f_{\rm incl}=0.7$), protecting them from destruction by SNe and thermal sputtering and reducing the amount of dust grain surface area available for accretion. 

We caution that the included physics in our dust evolution model are in no way complete and the predicted `critical' metallicities could change with the incorporation of new physics. However, the predictions of varying `critical' metallicities between dust species have some support from observations. 
As shown in Fig.~\ref{fig:sight_line_element_depletion_NH}, the LMC and SMC show large depletions of Si, Mg, and Fe in dense environments, suggesting the efficient growth of silicate and a theoretical iron-bearing dust species which agree with our model. While C depletions are currently unobservable in these galaxies, dust extinction and emission observations show decreasing amounts of small carbonaceous grains and PAHs \citep{weingartner_2001:DustGrainSizeDistributions,chastenet_2019:PolycyclicAromaticHydrocarbon}, suggesting carbonaceous dust growth may not be efficient in such galaxies. A sizable carbonaceous dust population dominated by large grains could exist and be indiscernible by current observational techniques. However, this seems unlikely as it would require carbonaceous grains to be far less prone to shattering into smaller grains compared to other dust species.

\subsection{Dust Buildup and Equilibrium} \label{sec:dust_buildup}

\begin{figure}
    \centering
    \includegraphics[width=0.99\columnwidth]{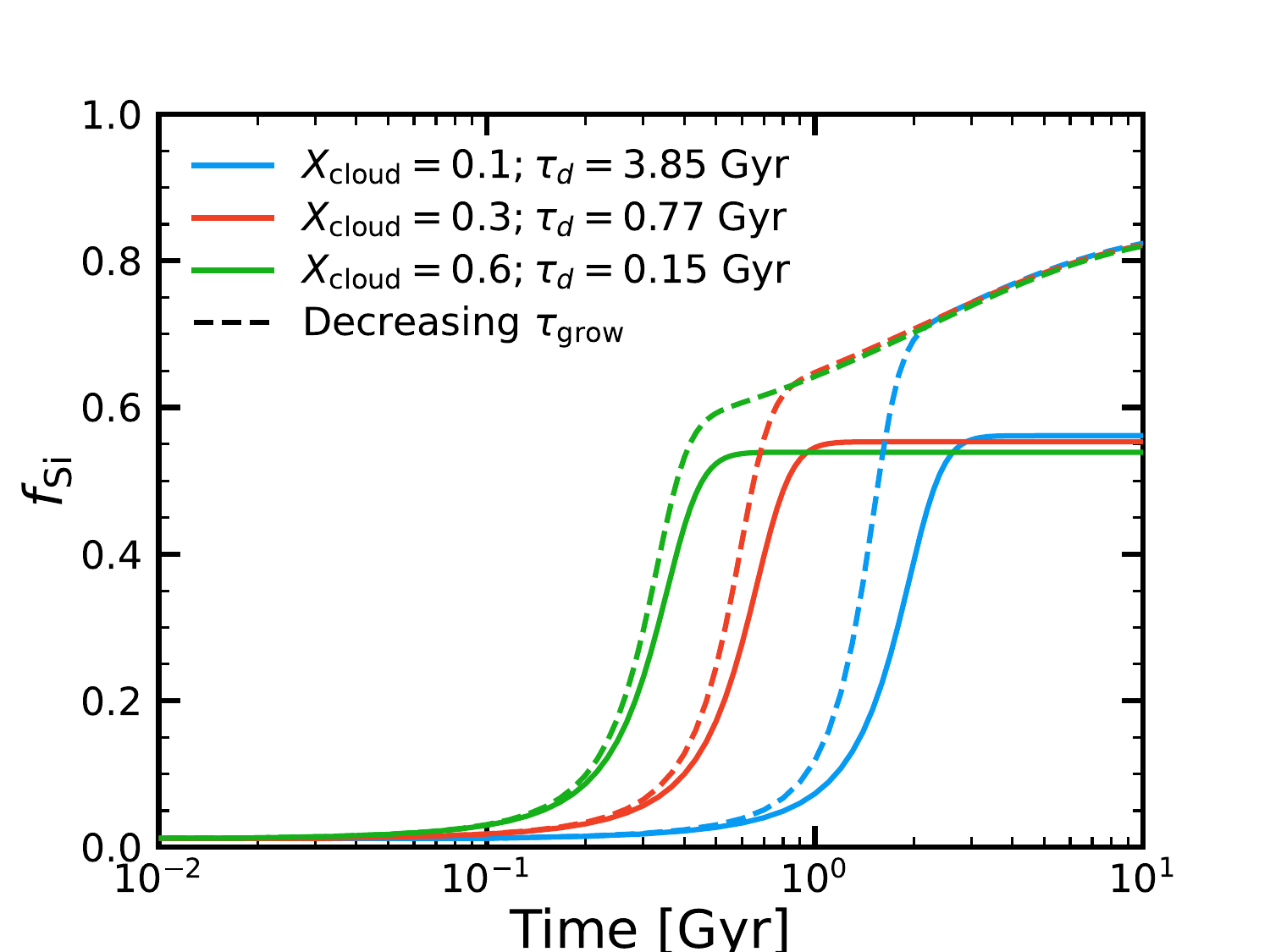}
    \caption{Predicted evolution of the fraction of Si locked in silicate dust through successive cycles of gas into dense clouds, where the dust grows via accretion, and out into diffuse clouds, where it is destroyed by SNe, using Eq. 20 and Eq. 21 in \citetalias{choban_2022:GalacticDustupModelling}. We assume an initial degree of condensation $f_{\rm Si}=0.01$, dust growth timescale $\tau_{\rm grow}\sim40$ Myr, and cold cloud lifetime $\tau_{\rm cloud}\sim10$ Myr ({\it solid}). We vary the SNe dust destruction timescale ($\tau_{\rm d}$) and the fraction of the ISM in cold clouds ($X_{\rm cloud}$), selecting values typical for ({\it red}) a MW-mass galaxy, ({\it blue}) a dwarf-mass galaxy, and ({\it green}) a massive/starburst galaxy.
    While the resulting equilibrium fractions are similar, the time to equilibrium varies dramatically $\teq\sim0.4-3$ Gyr, matching the spread in $\teq$ seen in our simulations. However, these predictions are an upper bound on $\teq$ since $\tau_{\rm grow}$ will decrease over time as the metallicity of a galaxy increases. We, therefore, also consider the case where $\tau_{\rm grow}$ decreases/metallicty increases ({\it dashed}). We assume the galactic metallicity has the general form $Z \sim A-B\exp^{-t / \tau}$, choosing $A$ and $B$ such that $\tau_{\rm grow,o}=40$ Myr initially and $\tau_{\rm grow,\infty}=10$ Myr at $t=\infty$, and $\tau=5$ Gyr. This modification shortens $\teq$ for each typical galaxy, but the variability remains large with $\teq\sim0.3-2$ Gyr.
    }
    \label{fig:equil_timescale}
\end{figure}

Once gas-dust accretion becomes efficient for a given dust species, the dust mass will build through successive cycles of gas into and out of dense environments until it reaches an equilibrium between dust growth and dust destruction via SNe.
This can be seen in Fig.~\ref{fig:all_galaxy_evolution},~\ref{fig:subsample1_dust_evolution}, \&~\ref{fig:subsample3_dust_evolution} as sharp increases in species mass fraction that eventually plateau (a similar but weaker trend can also be seen in D/Z). We label the time between the onset of gas-dust accretion and the equilibrium plateau as the `equilibrium' timescale ($\teq$) and find this timescale varies between dust species and between galaxies, with $\teq^{\rm iron}\sim0.75-4$ Gyr, $\teq^{\rm sil}\sim1.0-1.5$ Gyr, and $\teq^{\rm carb}\gtrsim3$ Gyr for metallic iron, silicates, and carbonaceous dust respectively. The variation in $\teq$ between dust species is due to the differences in dust life cycle processes outlined in Sec.~\ref{sec:efficient_accretion}, while the range in $\teq$ for a given dust species is due to differences in gas cycling and SNe destruction timescales between galaxies. 
We can confirm this with predictions of $\teq$ from analytical models (Eq. 20 and Eq. 21 in \citetalias{choban_2022:GalacticDustupModelling}).
In Fig.~\ref{fig:equil_timescale}, we show the predicted evolution of the fraction of Si locked in silicate dust
($f_{\rm Si}$) from this model. We assume an initial degree of condensation of $f_{\rm o,Si}=0.01$, cold cloud lifetime of $\tau_{\rm cloud}\sim10$ Myr estimated from Milky Way-mass galaxy simulations in \citet{benincasa_2020:LiveFastYoung}, and a constant accretion growth timescale of $\tau_{\rm grow}\sim40$ Myr which is the median value for gas in which accretion can occur in our model (T$<$300 K) for all galaxies in our sample at the onset of efficient silicate dust growth. We vary the SNe dust destruction timescale ($\tau_{\rm d}$; Eq. 18 in \citealt{mckee_1989:DustDestructionInterstellar}) and the fraction of the ISM with gas $T<300$ K ($X_{\rm cloud}$), which determines the average time it takes to cycle all ISM material from the cold cloud phase through the diffuse/warm ISM phases and back into cold clouds $\tau_{\rm cycle}=\tau_{\rm cloud} \frac{1-X_{\rm cloud}}{X_{\rm cloud}}$, to match typical values for dwarf, Milky Way, and massive/starburst galaxies.
By varying these two parameters, we predict a spread in $\teq$ similar to those exhibited in our simulations. However, these predictions are an upper bound on $\teq$, especially for galaxies with high star formation rates, since $\tau_{\rm grow}$ inversely scales with metallicity and will thus decrease over time.
We, therefore, also consider a simple case of decreasing $\tau_{\rm grow}$/increasing metallicity. If we assume the metal mass of a galaxy only changes due to metal injection from SNe and metal depletion via astration (metal locked up in stars) and a constant galactic gas mass, with no inflows or outflows, then the galactic metallicity will have the general form $Z\sim A - B\exp^{-t/\tau}$, where $A$, $B$, and $\tau$ are constants that depend on the SNe metal mass yields, star formation rate, and galactic gas mass. We show the results for the expected $\teq$ with this decreasing $\tau_{\rm grow}$ in Fig.~\ref{fig:equil_timescale} choosing values of $A$ and $B$ such that the growth timescales at $t=0$ and $t=\infty$ are $\tau_{\rm grow,o}=40$ Myr and  $\tau_{\rm grow,\infty}=10$ Myr respectively, and $\tau=5$ Gyr. Overall, this inclusion only slightly decreases the variation in $\teq$ but still overlaps with our simulation predictions. We caution that our simulations' resulting equilibrium timescales depend on our model's prescriptions for accretion and SNe dust destruction. In particular, our accretion routine assumes a constant size distribution, which will underpredict accretion timescales, and the theoretical predictions for SNe dust destruction efficiency vary substantially \citep[e.g.][]{hu_2019:ThermalNonthermalDust,kirchschlager_2019:DustSurvivalRates}.

The equilibrium timescale has important implications for the expected relation between galactic D/Z and metallicity when it is compared to the galactic metal enrichment timescale, $\tmetal$, which we define as the time it takes for the galactic metallicity to double.
For $\teq \lesssim \tmetal$, the dust population within a galaxy is able to reach a relative equilibrium D/Z for a given metallicity before it changes appreciably. This is similar to the model predictions presented in \citet{feldmann_2015:EquilibriumViewDust}, which assumes equilibrium galactic metallicity and D/Z. When $\teq \gtrsim \tmetal$, the dust population is still in the process of building up its mass and will `lag' behind the predicted equilibrium D/Z at a given metallicity. However, since $\teq$ inversely scales with metallicity, as the galactic metallicity increases, $\teq$ will eventually fall below $\tmetal$, causing these two trends to converge.
This effect can be seen in the evolution between galactic D/Z and Z for each galaxy shown in Fig.~\ref{fig:galaxy_int_DZ}, with the emergence of two evolutionary states we call `equilibrium' and `lagging'. 
The `equilibrium' state is exhibited by galaxies that have long quiescent periods, evolving in isolation for the majority of their life and/or experiencing relatively constant and low star formation rates (i.e. {\bf m11v\_halo2}, {\bf m11v\_halo0}, and {\bf m11i}) resulting in large $\tmetal$.
These galaxies all follow the same relation between D/Z and metallicity, similar to the equilibrium predictions from \citet{feldmann_2015:EquilibriumViewDust}.
The `lagging' state is exhibited by galaxies that have bursty episodes, experiencing major and/or multiple minor merger events and rapid changes in star formation rates (i.e. {\bf m12i}, {\bf m11d}, and {\bf m11e}) resulting in instances of small $\tmetal$.
These galaxies exhibit relations between D/Z and metallicity that occasionally fall below the `equilibrium' trend, which roughly corresponds with recent bursty periods.
These results suggest that recent galactic history plays a critical role in the expected galactic D/Z for galaxies in which gas-dust accretion is efficient and could explain a portion of the scatter in observed D/Z at a fixed metallicity.
However, our sample of galaxies is small, and the one galaxy that exhibits the largest deviation from equilibrium predictions is the only Milky Way-mass galaxy in our sample. More simulations are therefore needed to make a more statistically robust conclusion.

\section{Conclusions} \label{sec:conclusions}

In this work, we investigate the evolution of galactic dust populations, in both composition and amount, across cosmic time utilizing a suite of 7 cosmological zoom-in simulations run with the FIRE-2 model \citep{hopkins_2018:FIRE2SimulationsPhysics} for stellar feedback and ISM physics and the ``Species'' dust evolution model \citepalias{choban_2022:GalacticDustupModelling}. This dust evolution model accounts for dust creation in stellar outflows, growth from gas-phase accretion, destruction from SNe shocks, thermal sputtering, and astration, and turbulent dust and metal diffusion in gas. It tracks the evolution of specific dust species (silicates, carbon, silicon carbide), treating each uniquely depending on their chemical composition, along with theoretical nano-particle metallic iron (Nano-iron) dust species and an oxygen-based (O-reservoir) dust species. It also incorporates a physically motivated dust growth routine which accounts for Coulomb enhancement and CO formation in dense molecular environments. 

The 7 galaxies we selected cover a broad range of stellar ($10^6 \, \Msol < M_{*} < 10^{11} \, \Msol$) and halo ($10^{9} \, \Msol < M_{\rm vir} < 10^{12} \, \Msol$) masses at present day and showcase a variety of growth histories. We summarize our findings of dust population evolution below:

\begin{enumerate}

    \item Despite the variety of galactic evolutionary histories probed in our sample, all galactic dust populations follow similar evolutionary trends. Initially, they are dominated by dust production from first SNe II and later ($\sim2$ Gyr) AGB stars, resulting in a low D/Z $\sim0.01$. Above a `critical' metallicity threshold, gas-dust accretion becomes efficient and eventually increases the galactic D/Z $>0.1$ (Fig.~\ref{fig:all_galaxy_evolution},~\ref{fig:subsample1_dust_evolution}, \&~\ref{fig:subsample3_dust_evolution}). This suggests that gas-dust accretion is the main producer of dust mass for all but the most metal-poor galaxies and, in the case of Milky Way-mass galaxies, for a majority of the galaxy's life. 

    \item Differences between key element abundances, physical properties, and life cycle processes between dust species result in varying `critical' metallicities of $\Zcrit\sim0.05\Zsol$, $0.2\Zsol$, and $0.5\Zsol$ for metallic iron, silicates, and carbonaceous dust respectively. These differences reproduce observed depletions of Si, Mg, and Fe in the MW, LMC, and SMC (Fig.~\ref{fig:sight_line_element_depletion_NH} \&~\ref{fig:element_depletion_nH}), and suggest that silicate and a theoretical iron-bearing dust species grow efficiently by accretion in the LMC and SMC. They also suggest that C depletion decreases rapidly with galactic metallicity, which could explain the reduced amount of small carbonaceous grains observed in the LMC and SMC.

    \item After the onset of efficient gas-dust accretion, there is a characteristic equilibrium timescale over which dust mass builds up over time until an equilibrium between dust growth and SNe dust destruction is reached (Fig.~\ref{fig:equil_timescale}). This equilibrium timescale is $\gtrsim1$ Gyr in our sample and varies between galaxies. For galaxies with quiescent histories (long metal enrichment timescales), their dust populations are almost always in relative equilibrium and produce similar galaxy-integrated D/Z-Z trends. For galaxies with bursty episodes due to mergers/rapid changes in SF (short metal enrichment timescales), their dust populations can `lag' behind the expected equilibrium values for a time, producing lower D/Z. This `lagging' state can explain part of the large scatter in the observed relation between galaxy-integrated D/Z and metallicity (Fig.~\ref{fig:galaxy_int_DZ}).
    
    \item Extragalactic observations of spatially-resolved D/Z in spiral galaxies are roughly consistent with our Milky Way-mass galaxy (Fig.~\ref{fig:gal_resolved_DZ}). When compared to high-resolution observations of D/H$_{\rm neutral}$ in Local Group galaxies, our model overpredicts the amount of dust in diffuse neutral gas (Fig.~\ref{fig:DH_vs_SigmaHn}), but this may be due to our approximations of observables. A more direct comparison with mock dust emission SEDs is needed for a stronger test of our model.
    
\end{enumerate}

\section*{Acknowledgements}
CC and DK were supported by NSF grant AST-2108324. 
KMS was supported by NSF grant AST-2108081.
Support for PFH was provided by NSF Research Grants 1911233, 20009234, 2108318, NSF CAREER grant 1455342, NASA grants 80NSSC18K0562, HST-AR-15800.
CAFG was supported by NSF through grants AST-2108230, AST-2307327, and CAREER award AST-1652522; by NASA through grants 17-ATP17-0067 and 21-ATP21-0036; by STScI through grant HST-GO-16730.016-A; and by CXO through grant TM2-23005X. 
We ran simulations using: the Extreme Science and Engineering Discovery Environment
(XSEDE), supported by NSF grant ACI-1548562; Frontera allocations AST21010 and AST20016, supported by the NSF and TACC; Triton Shared Computing Cluster (TSCC) at the San Diego
Supercomputer Center.
The data used in this work were, in part, hosted on facilities supported by the Scientific Computing Core at the Flatiron Institute, a division of the Simons Foundation. This work also made use of MATPLOTLIB \citep{hunter_2007:Matplotlib2DGraphics}, NUMPY \citep{harris_2020:ArrayProgrammingNumPy}, SCIPY \citep{virtanen_2020:SciPyFundamentalAlgorithms}, and NASA’s Astrophysics Data System.

\datastatement{The data supporting the plots within this article are available on reasonable request to the corresponding author. A public version of the GIZMO code is available at \gizmourl.}



\bibliographystyle{mnras}
\bibliography{references} 



\appendix

\section{Effects of ISM Evolution - A Comparison Between FIRE-2 and FIRE-3} \label{Appendix_FIRE3}

\begin{figure*}
    \plotsidesize{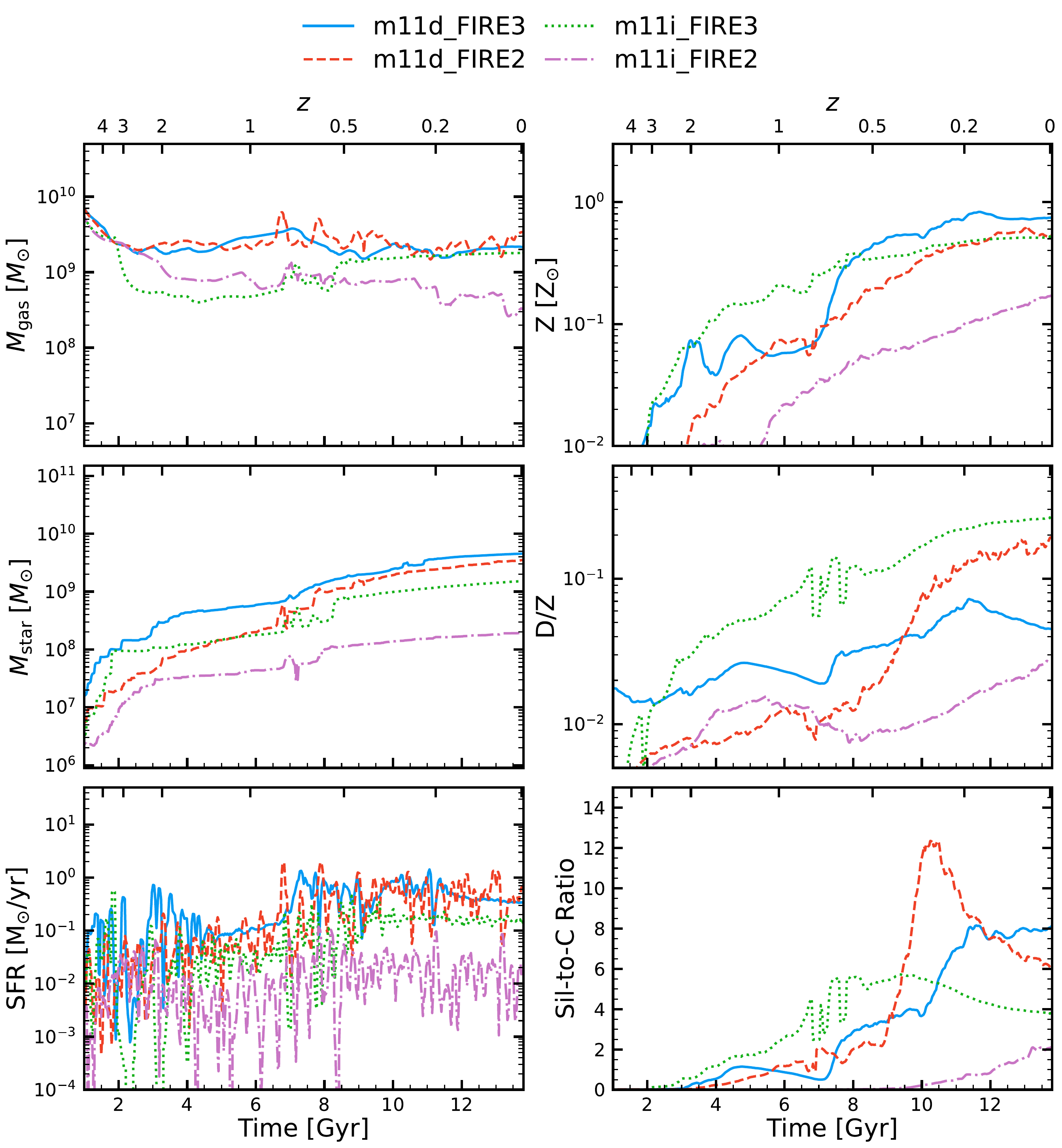}{0.8}
    \vspace{-0.25cm}
    \caption{Same as Fig.~\ref{fig:all_galaxy_evolution} for {\bf m11d} run with FIRE-3 ({\it solid}) and FIRE-2 ({\it dashed}) and {\bf m11i} run with FIRE-3 ({\it dotted}) and FIRE-2 ({\it dash-dotted}). Both galaxies experience higher star formation at early times and less bursty star formation for the later half of their lives with FIRE-3. The $z=0$ galactic properties for {\bf m11d} are similar between the different FIRE versions, but {\bf m11i} produces an order of magnitude more stars and a factor of 2 higher metallicity. However, this is most likely due to stochastic variations between simulation runs since {\bf m11i\_FIRE3} agrees with the original {\bf m11i} results presented in \citet{el-badry_2018:GasKinematicsMorphology} }  
    \label{fig:FIRE3_galaxy_evolution} 
\end{figure*}

\begin{figure*}
    \plotsidesize{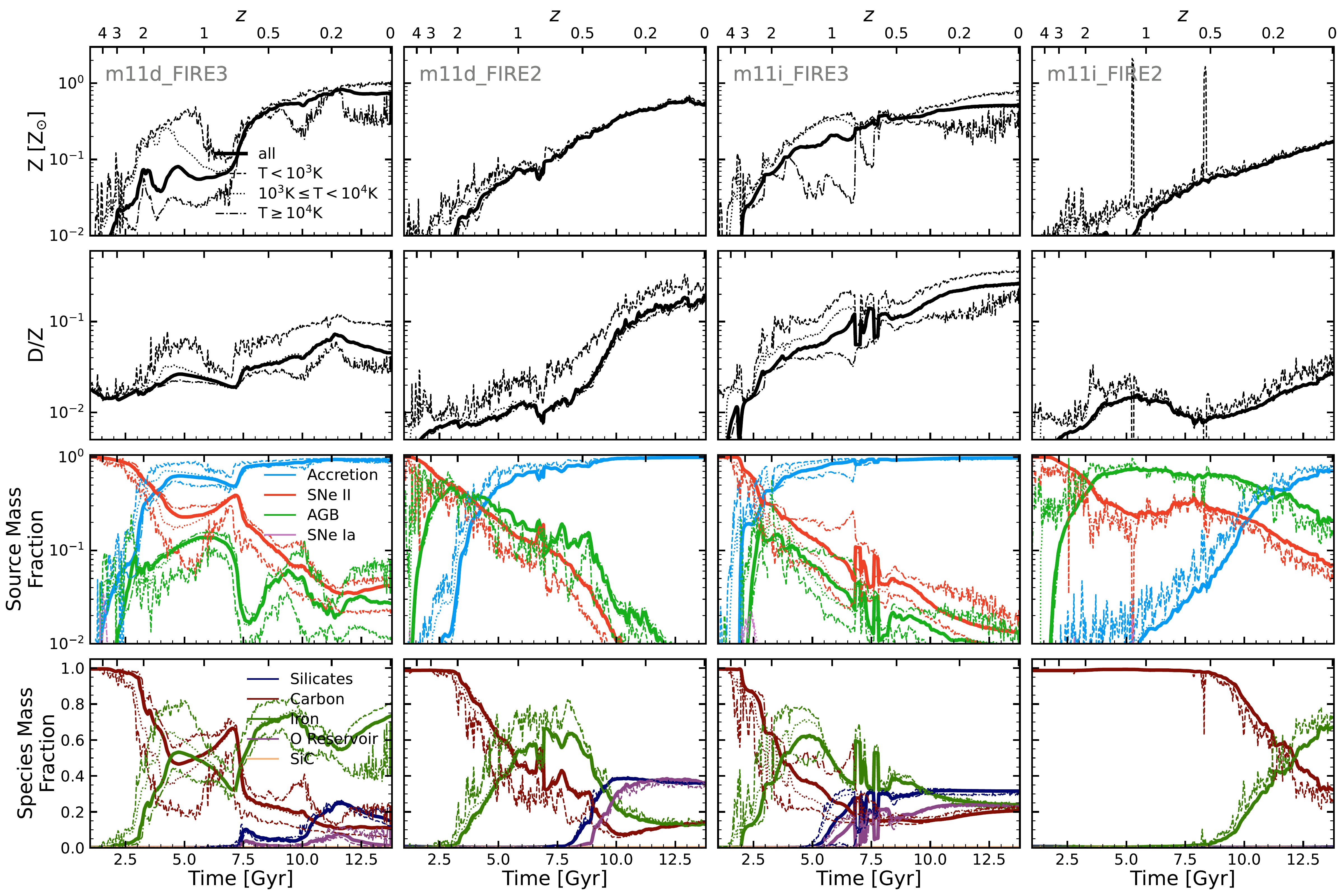}{0.99}
    \vspace{-0.25cm}
    \caption{Same as Fig.~\ref{fig:subsample1_dust_evolution} for {\bf m11d} run with FIRE-3 ({\it left}) and FIRE-2 ({\it left middle}) and {\bf m11i} run with FIRE-3 ({\it right middle}) and FIRE-2 ({\it right}). Regardless of which FIRE version is utilized, our main findings hold. Both galaxies exhibit `critical' metallicity thresholds for each dust species, above which the given dust species mass fraction rapidly increases due to efficient growth via accretion. The exact values of $\Zcrit$ vary between FIRE versions due to differences in adopted metal abundances and SNe metal yields. Both galaxies also exhibit a delay between the onset of efficient dust growth and when a given dust species reaches an equilibrium between growth and destruction. However, {\bf m11d\_FIRE3}'s dust population struggles to reach an equilibrium due to rapid changes in cold gas metallicity and its low cold gas fraction ($f_{\rm cold}\lesssim5\%$) compared to the other galaxies in our sample.}
    \label{fig:FIRE3_dust_evolution} 
\end{figure*}

All of our simulations in the main text are run with the FIRE-2 version of the FIRE code. The next version of FIRE, FIRE-3 \citep{hopkins_2023:FIRE3UpdatedStellar}, makes a variety of improvements to the stellar inputs and numerical methods, focusing in particular on updating the stellar evolution tracks used for stellar feedback and nucleosynthesis with newer, more detailed models, as well as improving the detailed thermochemistry of cold atomic and molecular gas, and adopting the newer \citet{asplund_2009:ChemicalCompositionSun} proto-solar reference abundances with $\Zsol\sim0.014$.
To test the robustness of our dust evolution results to these changes, we reran a subset of our simulation suite with FIRE-3. Specifically, we reran {\bf m11d} and {\bf m11i} (labeled {\bf m11d\_FIRE3} and {\bf m11i\_FIRE3}), with their resulting $z=0$ galactic properties given in Table~\ref{tab:FIRE3_simulations}. Fig.~\ref{fig:FIRE3_mockHubble} shows face-on mock Hubble {\it ugr} composite images for each galaxy at $z=0$. In  Fig.~\ref{fig:FIRE3_galaxy_evolution} \&~\ref{fig:FIRE3_dust_evolution}, we show the evolution of various galactic properties and a detailed breakdown of their metal and dust population evolution, respectively, both compared with the FIRE-2 runs.

While the exact differences in galactic evolution predicted by FIRE-2 versus FIRE-3 are beyond the scope of this paper, we will briefly comment on the broad differences exhibited by {\bf m11d\_FIRE3} and {\bf m11i\_FIRE3}. Both galaxies experience higher star formation at early times and less burst star formation at later times, resulting in a more prominent spiral shape at $z=0$ when compared to their FIRE-2 counterparts. While {\bf m11d\_FIRE3} has similar galactic properties at $z=0$, {\bf m11i\_FIRE3} produces an order of magnitude more stars, resulting in a $>2$ factor increase in metallicity.
However, this large difference is most likely due to stochastic variations between simulation runs since the resulting $z=0$ galactic properties of {\bf m11i} in our suite differ considerably from those in \citet{el-badry_2018:GasKinematicsMorphology}, where this simulation was first presented.

Looking at the resulting dust population evolution, these simulations agree with our main findings. Both galaxies exhibit `critical' metallicity thresholds as discussed in Sec.~\ref{sec:efficient_accretion} and predict that each dust species has a different $\Zcrit$. The exact $\Zcrit$ values differ slightly from the FIRE-2 predictions, but this is expected given the changes in reference element abundances (\citealt{asplund_2009:ChemicalCompositionSun} and \citealt{anders_1989:AbundancesElementsMeteoritic}) and SNe metal yields assumed in FIRE-2 and FIRE-3 which produce slight differences in key element abundances at a given metallicity. These simulations also exhibit an `equilibrium' timescale over which a dust species successively grows via accretion, as discussed in Sec.~\ref{sec:dust_buildup}. 
Similar to our FIRE-2 simulations, {\bf m11i\_FIRE3} exhibits species mass fractions that monotonically increase until an equilibrium is reached, producing $\teq\gtrsim1$ Gyr. On the other hand, {\bf m11d\_FIRE3} exhibits fluctuating species mass fractions that struggle to reach an equilibrium. Two factors cause this: {\bf (1)} the cold ($T<1000$ K) ISM gas metallicity varies considerably, as seen in Fig.~\ref{fig:FIRE3_dust_evolution}, moving above and below $\Zcrit$. {\bf (2)} {\bf m11d\_FIRE3} has an exceptionally low cold gas mass fraction ($f_{\rm cold}\lesssim5\%$) compared to {\bf m11i\_FIRE3} and {\bf m11d} ($f_{\rm cold}\gtrsim10\%$), resulting in long $\teq$.

\begin{table*}
	\centering
	\begin{tabular}{cccccccc} 
		\hline
		Name & $\Mvir \; (\msun)$ & $\Rvir$ (kpc) & $M_{*} \; (\msun)$ & $R_{1/2}$ (kpc) & Resolution ($\msun$) & Notes\\
		\hline
		m11d\_FIRE3 & 2.4E11 & 130 & 4.3E9 & 1.9 & 7100 & LMC-mass spiral w/ several mergers before $z\sim1$\\
            m11i\_FIRE3 & 6.4E10 & 84.1 & 1.4E9 & 2.6 & 7100 & LMC-mass spiral w/ major merger $z=0.8-0.5$\\
	   \hline
	\end{tabular}
	\caption{Same as Table~\ref{tab:simulations} for a subsuite of simulations rerun with FIRE-3 \citep{hopkins_2023:FIRE3UpdatedStellar} stellar feedback and ISM physics.}
    \label{tab:FIRE3_simulations}
\end{table*}

\begin{figure*}
    \plotsidesize{figures/FIRE3_Hubble.pdf}{0.9}
    \vspace{-0.25cm}
    \caption{Face-on mock Hubble images, same as Fig.~\ref{fig:mock_hubble_all}, of {\bf m11d} ({\it left}) and {\bf m11i} ({\it right}) rerun with FIRE-3 stellar feedback and ISM physics.} 
    \label{fig:FIRE3_mockHubble} 
\end{figure*}

\section{Sight Line Methodology} \label{Appendix_sightlines}

\begin{figure}
    \includegraphics[width=0.99\columnwidth]{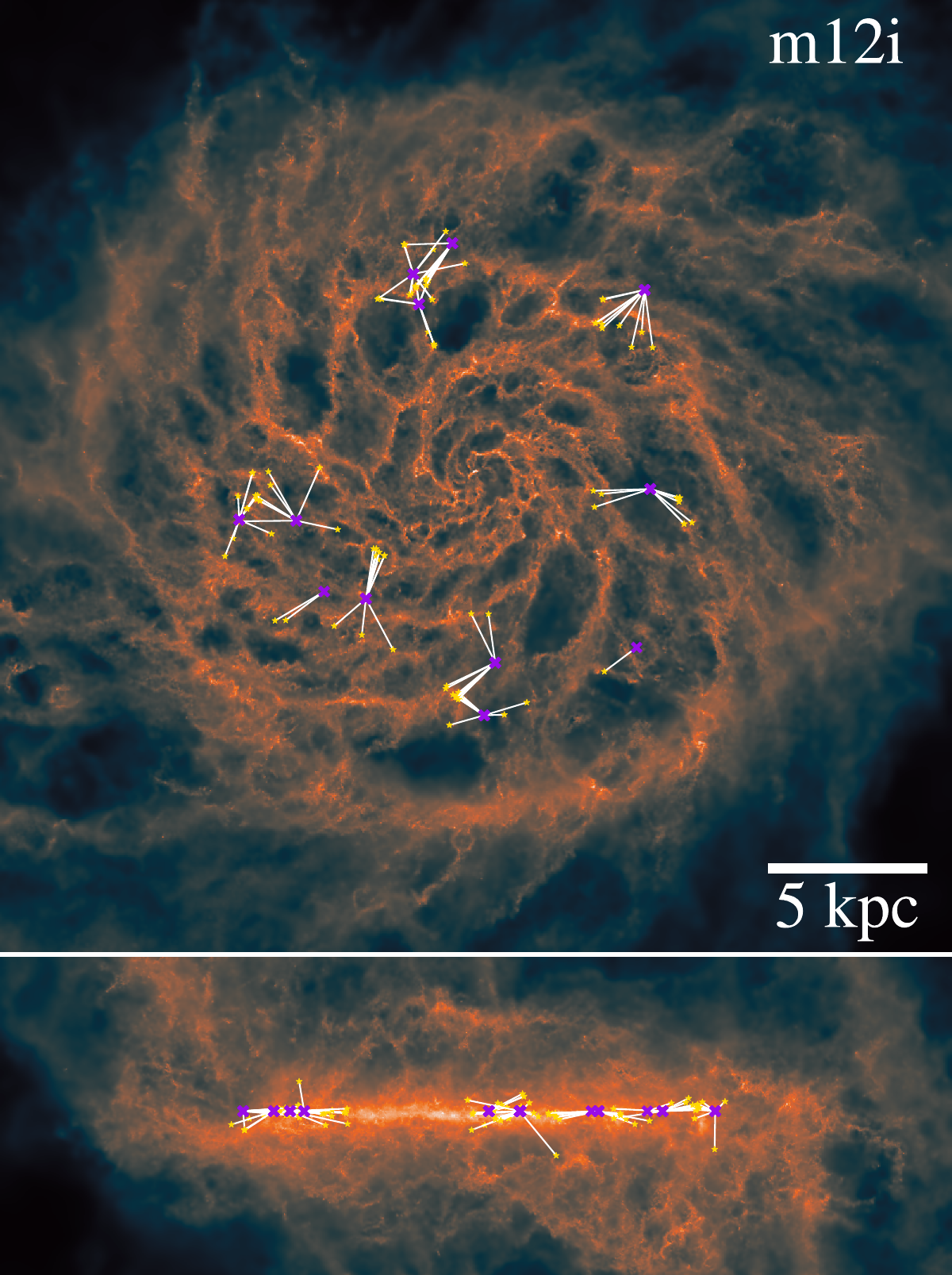}
    \vspace{-0.25cm}
    \caption{Face-on ({\it top}) and edge-on ({\it bottom}) gas density projection of {\bf m12i} with a subset of our sight lines ({\it white lines}) to young stars overplotted. The endpoints ({\it purple crosses}) of the sight lines are randomly selected within the disk. The start of the sight lines ({\it gold stars}) are up to 10 young star particles within $0.1-2$kpc of the endpoints. Our sight lines probe stars both within and above the disk plane, voids, and dense spiral arms.}  
    \label{fig:m12i_sightlines} 
\end{figure}

To compare directly to observations of element depletions in the MW, we created a set of $\sim$3000 sight lines within the galactic disk of {\bf m12i}. Each sight line is determined as follows: 
{\bf (1)} a point within an annulus ($r\sim5-8$kpc) of the galactic disk is chosen at random (this choice of range is due to {\bf m12i}'s less extended disk compared to the MW); this is the end of our sight line.
{\bf (2)} Up to 10 young star particles ($t_{\rm age}<10$ Myr) within $0.1-2$ kpc of the given point are chosen at random and will be the start of our sight lines. The specified stellar age and sight line distances are chosen to match the bias of observations to O \& B type stars and the given sight line distances from \citet{jenkins_2009:UnifiedRepresentationGasPhase}.
A gas density projection of {\bf m12i} with a subsample of sight lines overlaid is shown in Fig.~\ref{fig:m12i_sightlines}.

\begin{figure}
    \includegraphics[width=0.99\columnwidth]{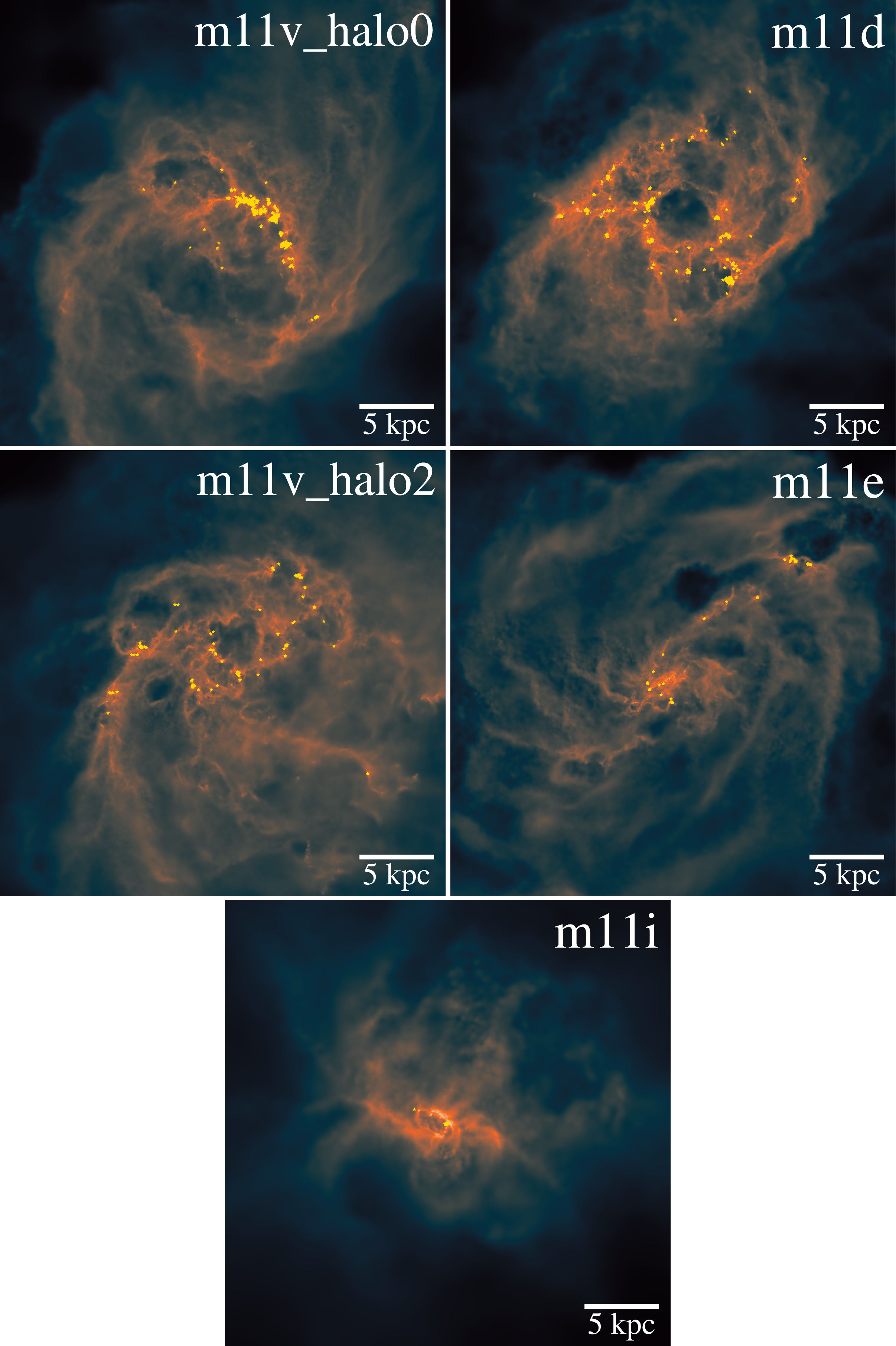}
    \vspace{-0.25cm}
    \caption{Face-on gas density projection of {\bf m11v\_halo0} ({\it top left}), {\bf m11d} ({\it top right}), {\bf m11e} ({\it middle left}), {\bf m11v\_halo2} ({\it middle right}), and {\bf m11i} ({\it bottom}) with all young stars ({\it gold stars}) overplotted. Three sight lines are created for each young star particle to simulate the view of an outside observer at different orientations. One is face-on with the average angular momentum vector of young stars in the galaxy, one is orthogonal to the angular momentum vector, and one is halfway between the two. Note that the number of young stars varies by up to 1 dex between the different galaxies however, they all roughly trace the densest parts of each galaxy.}
    \label{fig:dwarves_sightlines} 
\end{figure}

To compare directly to observations of element depletions in the LMC and SMC, we created a set of external sight lines for {\bf m11v\_halo0}, {\bf m11d}, {\bf m11e},{\bf m11v\_halo2}, and {\bf m11i}. Each sight line is determined as follows: 
{\bf (1)} The location of all young ($t_{\rm age}<10$Myr) star particles within 10 kpc of the galactic center are determined. This is the start of our sight lines. For reference, there are $\sim$1200, $\sim$600, $\sim$100, $\sim$200, and $\sim$30 young star particles within {\bf m11v\_halo0}, {\bf m11d}, {\bf m11e}, {\bf m11v\_halo2}, and {\bf m11i} respectively.
{\bf (2)} Three endpoints are determined to create three sight lines for each star particle. These endpoints are situated $\sim50$ kpc away (similar to the distance to the LMC and SMC) in three directions, one face-on with the average angular momentum vector of young stars in the galaxy, one orthogonal to the angular momentum vector, and one halfway between the two.
A gas density projection of the dwarf galaxies with young stellar populations overlaid is shown in Fig.~\ref{fig:dwarves_sightlines}.

We then use the {\small YTRay} object in {\small yt} \citep{turk_2011:YtMulticodeAnalysis} to
determine the gas cells intersected by a given sight line. This assumes spherical gas particles using the gas cell coordinates and smoothing kernel lengths from \GIZMO. Given a sight line start and end point, the intersected particles and their intersection lengths are determined. We then calculate the sight line's $N_{\rm X}$ and $\NHn$ from the sum of individual particle column densities determined from their H and element X number densities, assuming they are uniform within each cell, and intersection length.

\section{Defining Oxygen Abundance and Metallicity} \label{Appendix_OH}

\begin{figure*}
    \plotsidesize{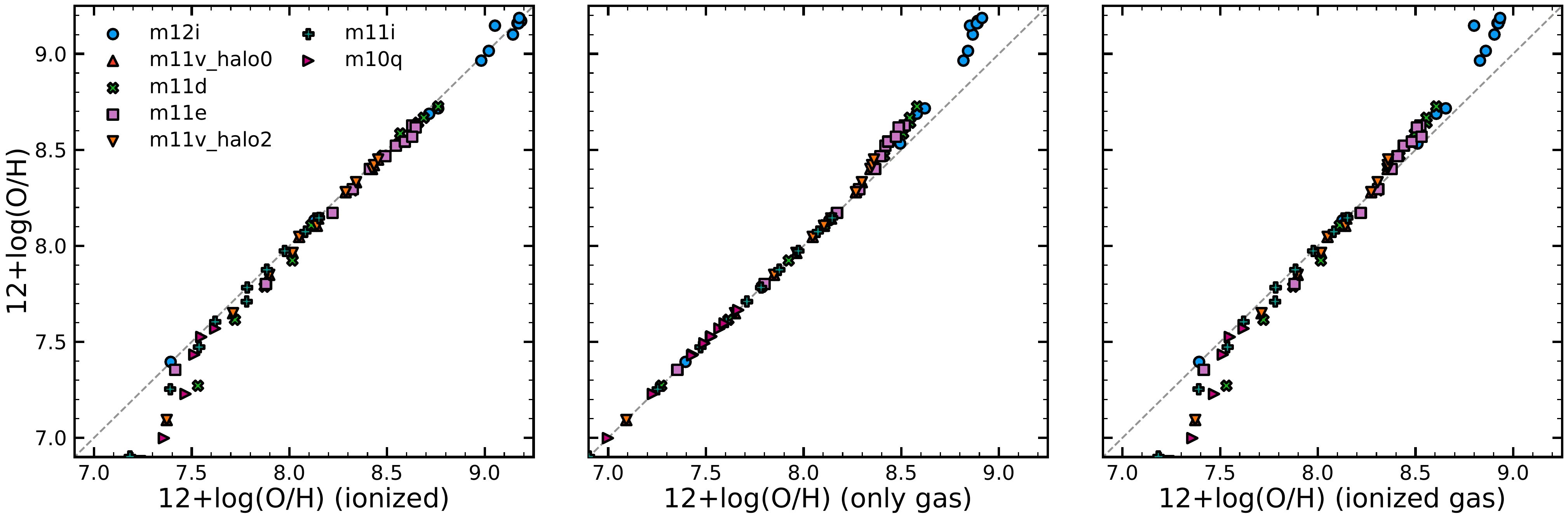}{0.99}
    \vspace{-0.25cm}
    \caption{Resulting oxygen abundances from different definitions for our suite of simulations. ({\it left}) Relation between oxygen abundances for definition (1), median O abundance for all gas within 10 kpc, and definition (2), median O abundance for all gas with $7000 < T < 15000$ K and $\nH > 0.5$ cm$^{-3}$. ({\it middle}) Relation between oxygen abundances for definition (1) and (3), median gas-phase O abundance (accounting for O depleted into dust) for all gas within $0.1 \Rvir$. ({\it right}) Relation between oxygen abundances for definition (1) and (4), median gas-phase O abundance (accounting for O depleted into dust) for all gas with $7000 < T < 15000$ K and $\nH > 0.5$ cm$^{-3}$. The data points shown are the same as those in Fig.~\ref{fig:galaxy_int_DZ}. Note the disagreement between definitions (1)  and (2) at low $\OH$ is due to our choice of only considering gas within 10 kpc of the galactic center. At early times, when $\OH$ is low, 10 kpc encompasses a large amount of ionized halo gas for our dwarf galaxies. If we only consider gas within $0.1\Rvir$ this disagreement disappears entirely.}
    \label{fig:O_definitions}
\end{figure*}

In this section, we investigate different definitions for galaxy-integrated oxygen abundance and discuss our choice of definition. These definitions are {\bf (1)} the mass-weighted median O abundance in gas cells with $7000 < T < 15000$ K and $\nH > 0.5$ cm$^{-3}$. This is motivated by galaxy-integrated observations which mainly probe auroral and nebular O emission lines in \textsc{Hii} regions \citep[i.e][]{pilyugin_2016:NewCalibrationsAbundance}. {\bf (2)} The mass-weighted median O abundance for all gas. {\bf (3)} \& {\bf (4)} The same as (1) and (2) but only gas-phase O abundance (accounting for O depleted onto dust). Note only some observations even account for possible depletion of metals into dust \citep[e.g][]{devis_2019:SystematicMetallicityStudy}.  
All of these definitions only consider gas within 10 kpc of the galactic center.
We show the differences between these definitions in Fig.~\ref{fig:O_definitions} for our suite of galaxies. On average, definitions (1) and (2) are consistent and match previous work from \citet[][Appendix A \& B]{ma_2016:OriginEvolutionGalaxy}\footnote{These definitions do diverge at low $\OH$, but this is due to our choice of only considering gas within 10 kpc, which is too large at early times/ low $\OH$. 10 kpc encompasses a large amount of ionized halo gas at these times, biasing definition (2) and (4). When we consider gas within $0.1 \Rvir$ this disagreement disappears entirely.}.
However, accounting for O depletion into dust causes the relation to diverge for $12+{\rm log_{10}(O/H)}>8.3$ due to the onset of efficient dust growth. This difference decreases for ionized gas due to the tendency for this gas to have experienced a recent SNe event, which destroys dust and reduces O depletion.

\begin{figure}
    \includegraphics[width=0.8\columnwidth]{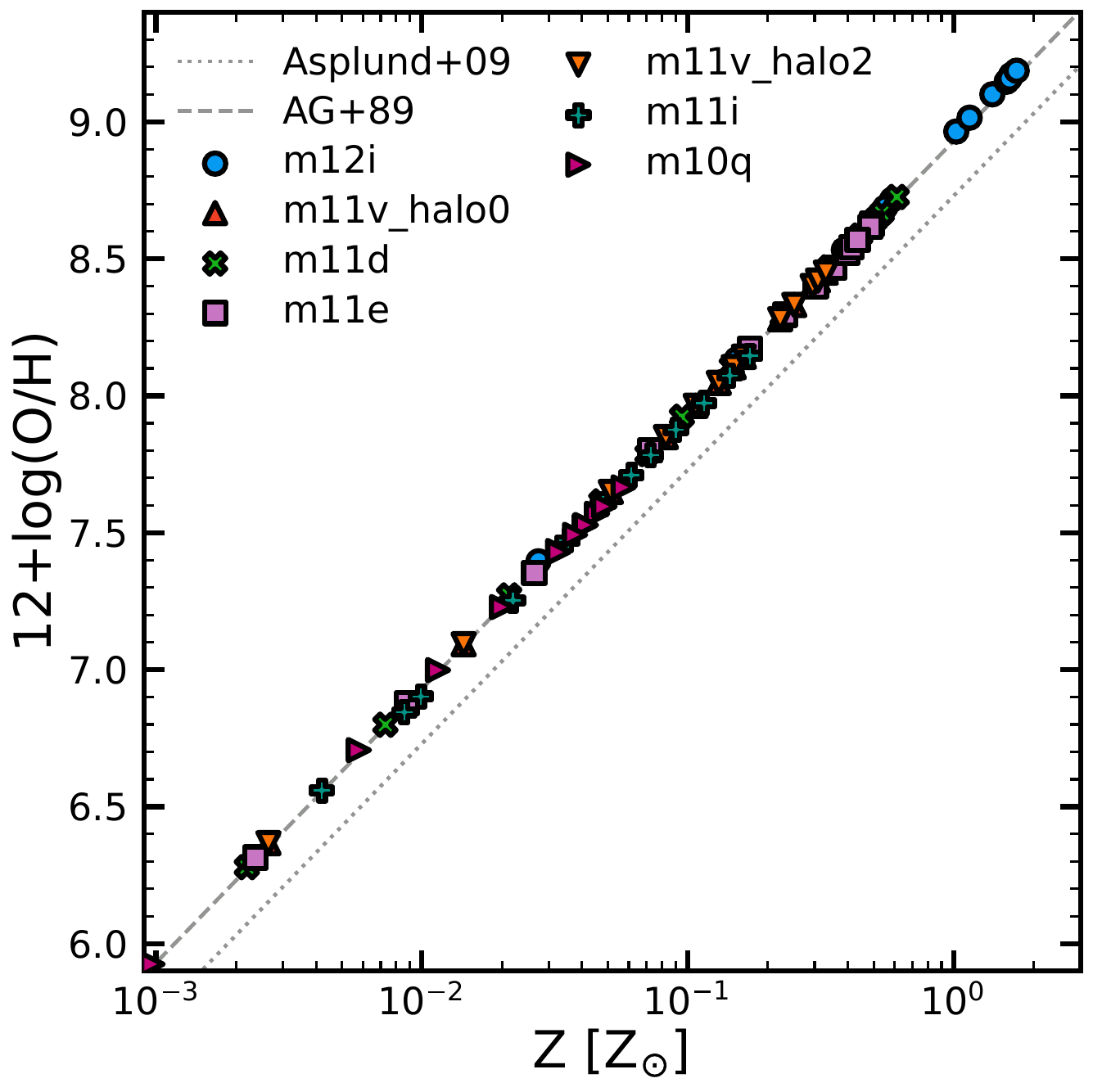}
    \vspace{-0.25cm}
    \caption{Resulting relation between median O abundance and median metallicity for all gas within 10 kpc for our suite of simulations. The data points shown are the same as those in Fig.~\ref{fig:galaxy_int_DZ}. Our simulations follow the O abundances expected from \citet{anders_1989:AbundancesElementsMeteoritic} which are offset by $\sim0.2$ dex from \citet{asplund_2009:ChemicalCompositionSun} abundances usually assumed by observations.}
    \label{fig:Z_vs_OH} 
\end{figure}

Another caveat is the assumed metal abundances in our simulations. FIRE-2 adopts abundances from \citet{anders_1989:AbundancesElementsMeteoritic} $(12+{\rm log_{10}(O/H)}_{\sun}=8.93)$, but most observations use more recent \citet{asplund_2009:ChemicalCompositionSun} proto-solar abundances $(12+{\rm log_{10}(O/H)}_{\sun}=8.73)$. We show the resulting relation between definition (1) and Z for our simulations in Fig.~\ref{fig:Z_vs_OH}. The median total O abundance closely follows the total metallicity of the galaxy and so including an overall offset of -0.2 dex in our definition of oxygen abundance will correct for this difference in adopted abundances.

For the reasons shown above, we define the galaxy-integrated $\OH$ as the weighted median gas phase O abundance (accounting for O locked in dust) in ionized regions, gas cells with $7000 < T < 15000$ K and $\nH > 0.5$ cm$^{-3}$, with a -0.2 dex offset.

\bsp	
\label{lastpage}
\end{document}